\documentclass{WileyMSP-template}

\begin{document}

%\pagestyle{fancy}
%\rhead{\includegraphics[width=2.5cm]{vch-logo.png}}
%\rhead{ }

\title{Nuclear Physics in the Era of Quantum Computing and Quantum Machine Learning}

\maketitle

% Author: Please give full first and last names for authors and include * after the name of all corresponding authors

\author{José-Enrique Garc\'{\i}a-Ramos*},
\author{Álvaro S\'aiz},
\author{José M. Arias},
\author{Lucas Lamata}, and
\author{Pedro Pérez-Fernández}

% Affiliations: Please provide adacemic titles (Prof. or Dr.) for all authors where applicable, and include an institutional email address for all corresponding authors

\begin{affiliations}
Dr. Jos\'e-Enrique Garc\'{\i}a-Ramos\\
Departamento de  Ciencias Integradas y Centro de Estudios Avanzados en F\'isica, Matem\'atica y Computaci\'on, Universidad de Huelva, 21071 Huelva, Spain\\
Instituto Carlos I de F\'{\i}sica Te\'orica y Computacional,  Universidad de Granada, Fuentenueva s/n, 18071 Granada, Spain\\ 
Email Address: enrique.ramos@dfaie.uhu.es\\ ~\\ 

\'Alvaro S\'aiz\\
Departamento de F\'{\i}sica Aplicada III, Escuela Técnica Superior de Ingeniería, Universidad de Sevilla, E-41092 Sevilla, Spain.\\
Email Address: asaiz@us.es\\ ~\\ 

Dr. Jos\'e M. Arias\\
Departamento de F\'{\i}sica At\'omica, Molecular y Nuclear, Facultad de F\'isica,  Universidad de Sevilla, Apar\-ta\-do 1065, E-41080 Sevilla, Spain\\
Instituto Carlos I de F\'{\i}sica Te\'orica y Computacional,  Universidad de Granada, Fuentenueva s/n, 18071 Granada, Spain\\
Email Address: ariasc@us.es\\ ~\\ 

Dr. Lucas Lamata\\
Departamento de F\'{\i}sica At\'omica, Molecular y Nuclear, Facultad de F\'isica, Universidad de Sevilla, Apar\-ta\-do 1065, E-41080 Sevilla, Spain\\ 
Instituto Carlos I de F\'{\i}sica Te\'orica y Computacional,  Universidad de Granada, Fuentenueva s/n, 18071 Granada, Spain\\
Email Address: llamata@us.es\\ ~\\ 

Dr. Pedro P\'erez-Fern\'andez\\ 
Departamento de F\'{\i}sica Aplicada III, Escuela Técnica Superior de Ingeniería, Universidad de Sevilla, E-41092 Sevilla, Spain.\\
Instituto Carlos I de F\'{\i}sica Te\'orica y Computacional,  Universidad de Granada, Fuentenueva s/n, 18071 Granada, Spain\\
Email Address: pedropf@us.es
\end{affiliations}

% Keywords: Please provide a minimum of three and a maximum of seven keywords, separated by commas


% Abstract should be written in the present tense and impersonal style (i.e., avoid we), and be at most 200 words long
\begin{abstract}
In this paper, the application of quantum simulations and quantum machine learning to solve low-energy nuclear physics problems is explored. The use of quantum computing to deal with nuclear physics problems is, in general, in its infancy and, in particular, the use of quantum machine learning in the realm of nuclear physics at low energy is almost nonexistent. We present here three specific examples where the use of quantum computing and quantum machine learning provides, or could provide in the future, a possible computational  advantage: i) the determination of the phase/shape in schematic nuclear models, ii) the calculation of the ground state energy of a nuclear shell model-type Hamiltonian and iii) the identification of particles or the determination of trajectories in nuclear physics experiments. 
\end{abstract}

\newpage

\section{Introduction}
\label{sec-intro}
In this perspective review, we discuss the link between low-energy nuclear physics and the emerging research field of quantum computing \cite{Niel2010}, which includes quantum simulations and quantum machine learning (QML) techniques. While both research fields have their own distinct problems and applications, they can be combined in a fruitful collaboration to yield new and relevant advancements. Although quantum simulations in nuclear physics have made progress in recent years, they have mainly focused on toy models or simple scenarios. However, the utilization of QML techniques in low energy nuclear physics is virtually non-existent. The objective of this perspective is to demonstrate the value of studying the combined fields of nuclear physics and quantum computing by providing a few examples and highlighting, to researchers in both domains, that this area of research holds great potential. As a matter of fact, in several long term plans and white papers these new research avenues have already been considered and promoted, namely, \cite{Carl2018} and \cite{Cloet2019} (US Department of Energy), the white paper \cite{Humb2022}, \cite{Beck2023} (US Department of Energy, National Science Foundation and National Institute of Standards and Technology) or the NuPECC Long Range Plan 2024 (still under discussion).
\bigskip

The structure of this work is as follows: Section \ref{sec-nuclear} revises the fundamentals of nuclear physics and presents its potential connections with quantum computing and QML. Next, in Section \ref{sec-QML}, we provide a brief overview of quantum simulations and QML. Section \ref{sec-cases} explores the current connections between quantum simulations, QML, and nuclear physics, illustrating this through a few examples such as: i) determining the shape/phase of a nucleus using the time evolution of an appropriated observable, ii) calculating the ground state energy of nuclei, and iii) identifying particles and reconstructing particle trajectories. Finally, in Section \ref{sec-conclusions}, we present our conclusions and provide an outlook for future research.

\section{The Nuclear Physics Realm}
\label{sec-nuclear}
In this section, we present in an abridged way, first, the most widely used nuclear physics models intended for low energy nuclear physics, second, the main lines of research of nuclear physics in quantum computing and, third, the most up to date applications of machine learning (ML) to treat nuclear physics problems. 

\subsection{Nuclear Models}
\label{sec-nuclear-models}
The study of the atomic nucleus is difficult since it is a quantum many-body system where two types of nucleons, protons and neutrons, interact via a force that is not completely known. In addition, the number of particles is not so large as to be appropriated the use of the powerful statistical mechanic machinery \cite{He2020BI}, therefore, it is necessary to consider a large number of degrees of freedom explicitly. Because of that, to understand the nuclear structure, one has to rely on nuclear models which can only describe partially the nuclear degrees of freedom. The situation is radically different from the studies in quantum chemistry, where the interaction is of Coulomb nature, therefore, fully known, %and a good agreement between theory and experiment is expected, 
although still a large number of particles should be considered. In the atomic nucleus a natural center for the potential does not exists, as it is the case of the atom, and the field that nucleons feel is created by the own nucleons. Hence, it is not obvious whether a  mean field could be defined \cite{He2020BI}. The theoretical description of the nuclear structure at low energy is based on three main approaches: i) the microscopic approach whose basic realization is the shell model \cite{He2020BI,Ta1993SM,RS2004MB}, ii) the mean-field approach \cite{RS2004MB,Niksic:2011sg,Grasso:2018pen, RRR-2018-rev}, and iii) the macroscopic approach based on the liquid drop model \cite{RS2004MB,BM1975nuclear,rowe2010nuclear}. %Within this last model the most important idea is the concept of nuclear surface and its deformation. %There are two competing interactions in nuclei: the pairing interaction that favors spherical shapes and the quadrupole-quadrupole interaction that produces deformed shapes: 
\begin{itemize}
 \item The nuclear shell model. It describes the behavior of nucleons (protons and neutrons) in an atomic nucleus. It is based on the independent particle motion motion of nucleons, being possible to define single particle orbits and, therefore, single particle levels, that occupy the nucleons within the nucleus, in a similar manner to electrons in an atom, i.e. in the atomic shell model. The model assumes that nucleons move in a mean field created by the whole set of nucleons and exhibit a quantum behavior. Once the nucleons are distributed in shells, the residual interaction between them is taken into account and the full diagonalization of the Hamiltonian is needed \cite{Ta1993SM}. The shell model successfully explains many nuclear properties, such as nuclear stability, magic numbers (nuclei with particularly stable configurations), nuclear spectra, and nuclear reactions. 
 %More recently, the Interacting Boson Model was introduced as a approximation to the Shell Model reducing the degrees of freedom to study the low-lying spectra of medium mass nuclei. The Interacting Boson Model (IBM) \cite{iachello1987interacting} is a nuclear model that describes the behavior of nuclei by considering that the most important degrees of freedom correspond to pairs of nucleons coupled either to angular momentum zero or two and that can be treated as bosons. The model possesses a u(6) dynamical algebra and all the operators of the model can be written in terms of the generators of that algebra. The IBM successfully predicts various nuclear properties, such as energy levels, electromagnetic transitions, and collective excitations, particularly in even-even nuclei. An additional advantage of the model is that it presents three dynamical symmetries that have been experimentally identified and that present closed formulas for the excitation energies and the transition probabilities. 
 \item The mean field, the beyond mean field approximation and the use of energy functional theories \cite{RS2004MB,Niksic:2011sg,Grasso:2018pen, RRR-2018-rev}. These rely on the assumption that nucleons move independently in an effective average potential generated by all other nucleons. The interaction can be obtained globally in a self-consistent way for the whole mass table, fixing a set of free coefficients to reproduce ground state properties of all known nuclei. This simplification allows for the treatment of complex many-body systems by reducing the problem to an effective single-particle problem. Theoretical foundations of mean-field models include: i) the Hartree-Fock or Hartree-Fock-Bogoliubov formalism, ii) density functional theory (DFT), and iii) mean-field potentials and self-consistency. Applications of mean-field models in nuclear structure include the calculations of nuclear binding energies and masses, nuclear deformation and shape transitions, shell structure and magic numbers, and collective motion and excitations, among others.

 \item  The collective model. Also known as the liquid drop model, it treats the nucleus as a droplet of incompressible nuclear matter. This model assumes that the nucleus behaves like a classical liquid drop, with nucleons interacting through attractive and repulsive forces. It explains nuclear phenomena by considering collective motion, such as rotation and vibration, of the nuclear surface. The model successfully describes phenomena like nuclear deformation, fission, and certain aspects of nuclear spectra. Along the same lines, the Kumar-Baranger model \cite{KB1,KB2,KB3}, or the generalized collective model, is an extension of the collective Bohr model \cite{Bohr1998}. It takes into account the coupling between collective motion and single-particle excitations within the nucleus. This model considers both vibrational and rotational degrees of freedom and is especially useful for describing transitional nuclei that exhibit characteristics of both vibrational and rotational motion.
\end{itemize}

These models, and their extensions, have significantly contributed to our understanding of nuclear structure. The difficulty in solving the nuclear physics problem, apart from the related ones with the interaction, is the dimension of the Hilbert space that for medium-mass and heavy nuclei around the center of a major shell, either in protons, neutrons or both, is far beyond the present and even future computational capabilities. In the case of the shell model, an explosion of the dimension of the Hilbert space appears when a large number of nucleons should be distributed in large major shells. Moreover, in order to explain certain phenomena it is needed to allow multi particle-hole excitations across two major shells, which generates a further explosion of the Hilbert space dimension. In the case of the energy functional approach, the major issue is connected with the generation of states with defined quantum numbers, i.e., to go beyond mean field using, for instance, the generator coordinate method \cite{RS2004MB}, which involves the evaluation of integrals that, once more, can be computationally very costly. The use of quantum computing and, in particular, quantum machine learning can open a new avenue to deal with these nuclear problems in the near future.   
\bigskip

Nowadays, the implementation of state-of-the-art nuclear shell model or beyond mean-field problems in Noisy-Intermediate Scale Quantum (NISQ) computers is not yet possible, because of the number of available qubits, but also because of the necessity of developing new algorithms to implement the nuclear problem in the available nuclear hardware. Because of that, some simpler models that retain the main characteristics of the aforementioned ones have been recently used:  i) the Lipkin-Meshkov-Glick (LMG) model \cite{LMG} that is valid for representing particular nuclear systems but it is also of great interest in quantum optics and also describes, in an approximate way, certain solid state systems and ii) the Agassi model \cite{AGASSI196849} that mimics the interplay between pairing and quadrupole interactions as in the Kumar-Baranger model.

%\bigskip

%The goal of this perspective is to present a few selected topics in nuclear physics that could benefit or have already been implemented in the quantum computing and QML realms. The selected cases are: i) determination of the shape/phase of a nuclear system through its time evolution, ii) determination of the ground state in the nuclear shell model, and iii) identification of particles and reconstruction of trajectories in nuclear physics experiments. 

\subsection{Nuclear Physics and Quantum Computing}
\label{sec-nuclear-QC}
%Nuclear physics and quantum computing are two distinct fields that have their own areas of study and %applications. %However, there are some intersections between them:
%\begin{itemize}
% \item  
%    Quantum computing is a rapidly advancing field that utilizes the principles of quantum mechanics to perform computation. Instead of classical bits, which can be either 0 or 1, quantum computers use quantum bits or qubits, which can exist in superposition, representing both 0 and 1 simultaneously. Quantum computers leverage quantum phenomena such as entanglement and interference to perform certain computations exponentially faster than classical computers.
%\item
%    Nuclear Physics and Quantum Computing:
%    Nuclear physics, on the other hand, as mentioned above, is concerned with the study of atomic nuclei, their structure, and their interactions. It focuses on understanding the fundamental forces and particles that govern the behavior of atomic nuclei. 
%\end{itemize}

Nuclear physics and quantum computing are two rather distinct fields. However, there are some potential connections and applications where they intersect:
\begin{itemize}    
\item Quantum simulations: quantum computers could have the potential to simulate quantum systems more efficiently than classical computers. This includes simulating the behavior of atomic nuclei and nuclear interactions. By harnessing the power of quantum computing, researchers may be able to gain deeper insights into nuclear physics phenomena. 
\item Quantum algorithms for nuclear physics: quantum algorithms, such as the Variational Quantum Eigensolver (VQE) \cite{Peru014,Tilly2022} and Quantum Phase Estimation (QPE) \cite{Niel2010}, have been developed to solve problems in quantum chemistry, which has overlap with nuclear physics. These algorithms can be applied to nuclear physics problems, such as calculating nuclear energy levels or simulating nuclear reactions.
\item Data analysis and optimization: nuclear physics experiments generate vast amounts of data that need to be analyzed and optimized. Quantum computing techniques, such as QML algorithms or quantum optimization algorithms, may offer novel approaches to process and extract valuable insights from nuclear physics data.
\end{itemize}

While the full extent of the connection between nuclear physics and quantum computing is still being explored, it is an exciting area of potential collaboration and research. The development of more powerful and scalable quantum computers may provide new tools and techniques to address complex nuclear physics problems.

\subsection{Nuclear Physics and Machine Learning}
\label{sec-nuclear-ML}
In the last few years, ML has been applied to nuclear physics in different tasks \cite{Boeh2022}:
\begin{itemize}
\item Data analysis: nuclear physics generates large amounts of experimental data. ML can help to process and analyze this data efficiently to extract patterns, identify particles, perform classifications, and make predictions. ML algorithms such as neural networks can uncover hidden correlations and trends in nuclear data.
\item Nuclear modeling: ML can also be used to develop models and simulations in nuclear physics. Traditional nuclear physics models often involve complex calculations and approximations. ML offers alternative approaches to nuclear modeling, where ML algorithms can be used to construct empirical models from nuclear data and improve prediction accuracy.
\item Particle detection: in nuclear physics experiments, it is crucial to identify and track charged particles. ML has been successfully applied to particle detection and trajectory reconstruction, which can enhance the precision and efficiency of data analysis. Algorithms such as pattern classifiers, convolutional neural networks (CNN), and particle tracking algorithms can aid in particle identification and reconstruction in nuclear detectors.
\item Experimental optimization: ML can assist in the optimization of nuclear physics experiments. Optimization algorithms such as genetic algorithms or reinforcement learning can help finding optimal configurations of experimental parameters, thereby saving time and resources in data collection.
\end{itemize}

These are just a few areas where ML has been successfully applied in nuclear physics. The intersection of both disciplines offers exciting opportunities to improve data analysis, develop more accurate models, and optimize experiments. As ML continues to evolve, its application in nuclear physics is likely to expand further. 

\section{Quantum Simulations and Quantum Machine Learning}
\label{sec-QML}
Quantum simulations \cite{Geor2014} is a rapidly growing area of research in which a quantum controllable system is employed to reproduce the properties (either dynamical or static) of another quantum system of interest. Several proposals and experiments in this field have been produced in the past two decades, involving, as quantum simulator platforms, trapped ions, superconducting circuits, cold atoms, quantum photons, and nuclear magnetic resonance, among others. The simulated quantum systems are diverse and could be roughly grouped in condensed matter, quantum chemistry, and high-energy physics, although this is not an exclusive list \cite{Geor2014}. Inside the field of quantum simulations, and in the intersection of many-body quantum systems and high-energy physics, a fledgling field has appeared in the past few years \cite{Bauer2023}.

\bigskip 

Quantum simulators belong mainly to one of three possible categories: digital quantum simulators, analog quantum simulators, and digital-analog quantum simulators \cite{DAQS_Review}. Digital quantum simulators allow for a wide variety of systems to be simulated, as they have universality properties: they decompose the simulated quantum dynamics into elementary unitary gates, that later on are implemented in the quantum simulator in a successive way. This is done via a Lie-Trotter-Suzuki expansion. The main drawback of digital quantum simulators is that it is difficult to go beyond a few dozen qubits with current technology, due to the accumulated gate errors as the single- and two-qubit gates in which the protocol is decomposed are never perfect. Moreover, there is usually a digital error as well, due to the fact that the different gates do not commute with each other. 
\bigskip 

On the other hand, analog quantum simulators implement in the quantum platform a quantum dynamics following a Hamiltonian that is similar to the one of the simulated quantum system, by tuning some controls such as laser pulses, microwaves, etc. The advantage of analog quantum simulators is that they are more scalable than the purely digital ones as they have less accumulated gate error as well as digital error.
\bigskip

Finally, digital-analog quantum simulators aim at benefiting from both paradigms, digital and analog, via combining large analog blocks (which provide scalability) with digital steps (which enable to simulate a wider variety of models than the purely analog). This paradigm could be a way of achieving useful new knowledge in NISQ Computers in the near and mid term \cite{DAQS_Review}.
\bigskip

The typical errors in a quantum simulator can be the mentioned ones: accumulated gate error and digital error, as well as the common to all quantum systems, such as decoherence due to an uncontrollable coupling to the environment. Therefore, sometimes it is advisable to employ a master equation to theoretically model a quantum simulation platform, as well as to interpret the experimental results \cite{Geor2014}.
\bigskip

QML \cite{QML_Nature_Review,Lamata_2020,Lamata_2023}  aims at connecting the two timely fields of ML (in turn belonging to the more general artificial intelligence) and quantum computing. The goal is either to employ quantum devices to carry out more efficient ML calculations, or to use ML algorithms to better control and analyze quantum systems.
\bigskip

The motivation to explore this field is the fact that quantum mechanics is described by the formalism of linear algebra, which is a discipline that is also commonly employed in ML, e.g., for computing distances when classifying data. Thus, it is sensible to aim at using quantum devices to carry out some of the ML tasks, which suffer from the so-called dimensionality curse, far more efficiently. Namely, with less time and energy resources expense \cite{QML_Nature_Review,Lamata_2020,Lamata_2023}.
\bigskip

The field of QML has significantly grown in the past five years, and several theory proposals as well as experimental realizations have been produced, in platforms such as superconducting circuits, quantum photonics, and trapped ions \cite{QML_Nature_Review,Lamata_2020,Lamata_2023}. However, the use of QML protocols for the analysis of nuclear physics is a relatively unexplored field so far.
\begin{figure}
\begin{tabular}{cc}
    %\centering
    \vspace{0pt} \hspace{30pt} \textbf{(a)} & \hspace{20pt} \textbf{(b)} \hspace{100pt} \textbf{(c)} \\ 
    \vspace{5pt} \includegraphics[width=0.28\linewidth]{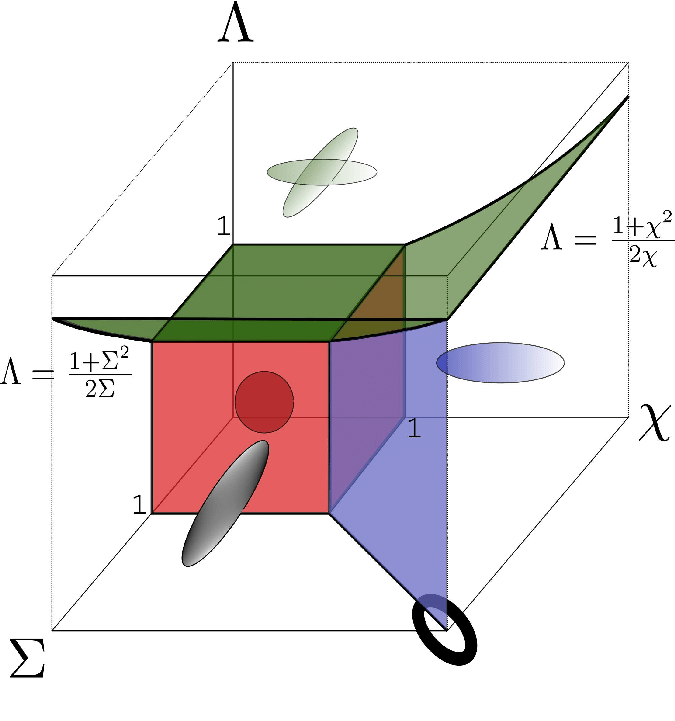} &
    %\centering
   \includegraphics[width=0.45\linewidth]{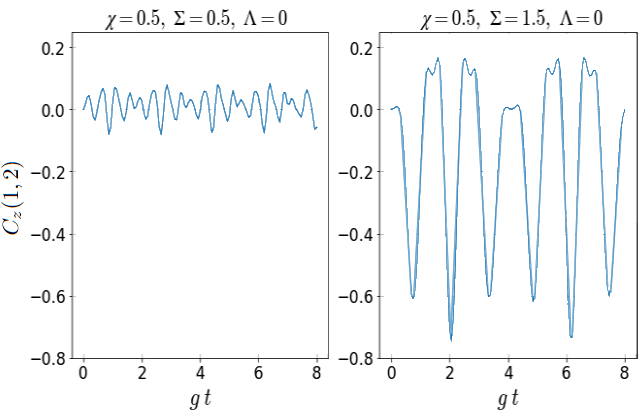}\\
    %%%%%%%
    %\hspace{30pt} d) \hspace{110pt} e)
    %\centering
    \hspace{20pt} \textbf{(d)} \hspace{100pt} \textbf{(e)} & \hspace{20pt} \textbf{(f)} \hspace{100pt} \textbf{(g)} \\
    \includegraphics[width=0.45\linewidth]{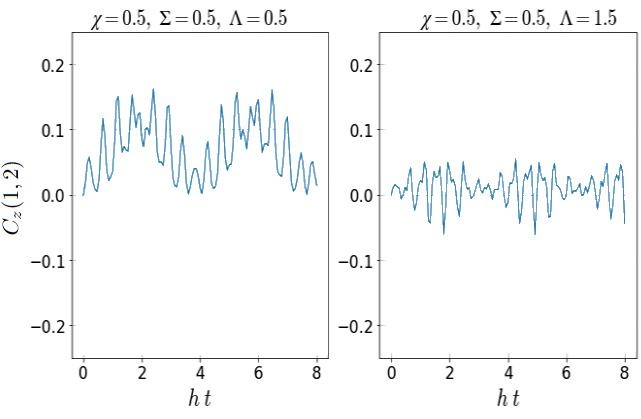}&
    %\hspace{30pt} f) \hspace{110pt} g)
    %\centering
    \includegraphics[width=0.45\linewidth]{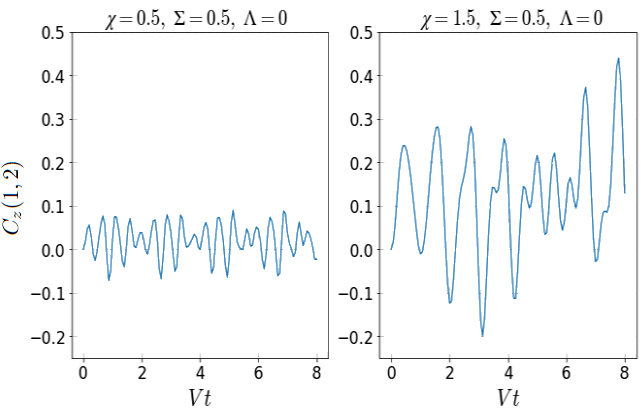}
\end{tabular}
\caption{Panel (a) Phase diagram of the extended Agassi Hamiltonian. For convenience the rescaled parameters $\chi$, $\Sigma$, and $\Lambda$ are used ($V = \frac{\varepsilon \chi}{2j - 1}$,  $g = \frac{\varepsilon \Sigma}{2j - 1}$, and  $h = \frac{\varepsilon \Lambda}{2j - 1}$ where $2 j$ corresponds to the size of the shell).  Red vertical planes represent second-order QPT surfaces. The green surface and the blue vertical one  correspond to first-order critical surfaces. The symbols represent the different possible phases of the system (see \cite{Garc2018}). 
%Red sphere, blue oval, black oval, black thick oval and crossed green ovals correspond to the symmetric solution, the HF deformed solution, the BCS deformed solution, the closed valley solution and the HF-BCS deformed solution, respectively.  
Panels (b)-(g) correspond to the time evolution of the $C_{z}(1,2)$ for selected values of the control parameters. Reproduced with permission \cite{Saiz2022}. Copyright 2022, American Physical Society.} 
\label{fig-phase-dia-time}
\end{figure}
\bigskip

Some of the quantum algorithms being employed in QML are quantum versions of standard ML ones, such as quantum supervised learning, quantum unsupervised learning, and quantum reinforcement learning. Quantum supervised learning employs a quantum algorithm for solving linear systems of equations, the Harrow-Hassidim-Lloyd (HHL) algorithm \cite{QML_Nature_Review}. Other kinds of algorithms, such as quantum reinforcement learning, employ in some cases the speedup of Grover search algorithm to improve the performance of standard reinforcement learning. Some more prominent QML algorithms that are emerging are the parameterized quantum circuits, also named quantum neural networks, as well as quantum kernels and quantum feature spaces  \cite{QML_Nature_Review,Lamata_2020,Lamata_2023}.

\section{Quantum Simulations and Quantum Machine Learning for Low Energy Nuclear Physics} 
\label{sec-cases}
\subsection{Determination of the Shape of a Nuclear System through its Time Evolution}
\label{sec-shape}
Quantum Phase Transitions (QPT's) appear in quantum systems at zero temperature when a sudden change in the ground-state structure appears under a change of a control parameter in the Hamiltonian \cite{Sach11}, changing, therefore, the shape of the system. A typical situation in which a QPT is present corresponds to Hamiltonians that can be written as two pieces with different symmetries (A and B):
\begin{equation}
\label{eq:H_QPT}
\hat H= (1-x) \hat H_A+ x\hat H_B.
\end{equation}
This formulation allows us to investigate the interplay between the two symmetries, A and B, by adjusting the control parameter $x$, which determines the relative contribution of each symmetry to the overall Hamiltonian. Under the formulation (\ref{eq:H_QPT}) two clear limiting situations exist: A for $x=0$, and B for $x=1$. These limits typically correspond to dynamical symmetries \cite{iachello1987interacting}. However, for $x-$values different from 0 and 1, the Hamiltonian has no definite symmetry and A and B compete among them. In spite of this lack of symmetry, interestingly enough the system remains close to A or B until the critical point, $x=x_c$, at which a sudden change in the system structure appears (QPT). The existence of a QPT also implies a sudden change in the so-called order parameter, which vanishes in one of the phases (symmetric) and takes a nonzero value in the other phase (broken or non-symmetric phase) \cite{Sach11}. QPTs can be classified accordingly to the Ehrenfest classification \cite{Land69} in a similar manner to the phase transitions that occur in macroscopic systems at non-zero temperature.
\bigskip

\begin{figure}
    \centering
    \includegraphics[scale=0.7]{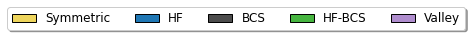}
    \begin{tabular}{c c}
        \vspace{3pt}
        \textbf{ (a)} $\Sigma = 0.5$  ;  $\Lambda = 0.5$ & \textbf{(b)} $\chi = 0.5$  ;  $\Lambda = 0.5$ \\
        \hspace{-6pt}\begin{tabular}{c c}
            \vspace{-5pt}
            Exact & Trotter \\
            \includegraphics[scale=0.5,trim={8pt 0 0 0},clip]{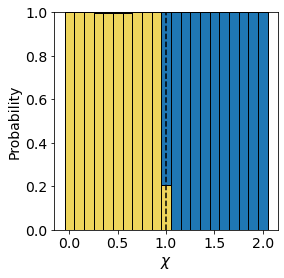} &
            \hspace{-5pt}\includegraphics[scale=0.5,trim={50pt 0 0 0},clip]{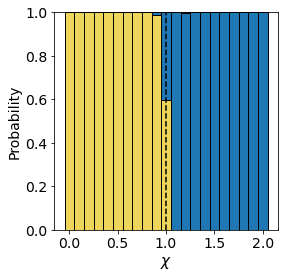}
        \end{tabular} &
        \hspace{-6pt}\begin{tabular}{c c}
            \vspace{-5pt}
            Exact & Trotter \\
            \includegraphics[scale=0.5,trim={8pt 0 0 0},clip]{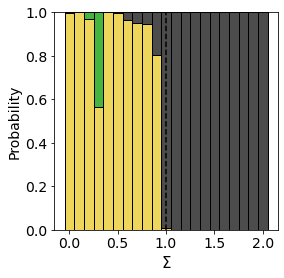} &
            \hspace{-5pt}\includegraphics[scale=0.5,trim={50pt 0 0 0},clip]{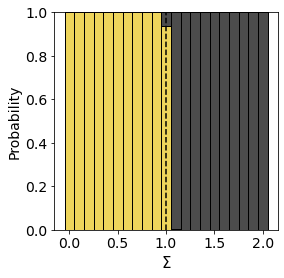}
        \end{tabular} \\
        \vspace{3pt}
        \textbf{(c)} $\chi = 0.5$  ;  $\Sigma = 0.5$ & \textbf{(d)} $\chi = 1.5$  ;  $\Lambda = 0.5$ \\
        \hspace{-6pt}\begin{tabular}{c c}
            \vspace{-5pt}
            Exact & Trotter \\
            \includegraphics[scale=0.5,trim={8pt 0 0 0},clip]{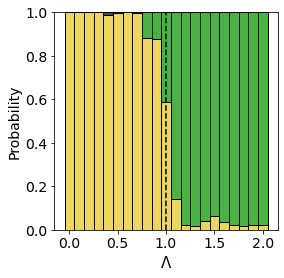} &
            \hspace{-5pt}\includegraphics[scale=0.5,trim={50pt 0 0 0},clip]{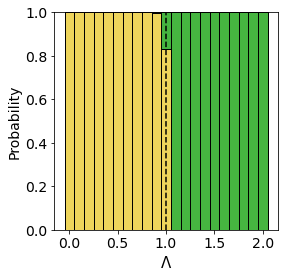}
        \end{tabular} &
        \hspace{-6pt}\begin{tabular}{c c}
            \vspace{-5pt}
            Exact & Trotter \\
            \includegraphics[scale=0.5,trim={8pt 0 0 0},clip]{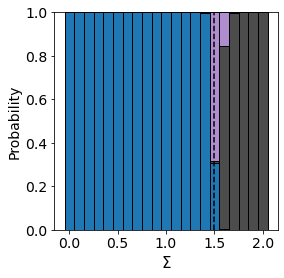} &
            \hspace{-5pt}\includegraphics[scale=0.5,trim={50pt 0 0 0},clip]{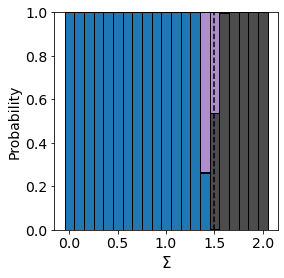}
        \end{tabular} \\
        \vspace{3pt}
        \textbf{(e)} $\chi = 1.5$  ;  $\Sigma = 0.5$ & \textbf{(f)} $\chi = 0.5$  ;  $\Sigma = 1.5$ \\
        \hspace{-6pt}\begin{tabular}{c c}
            \vspace{-5pt}
            Exact & Trotter \\
            \includegraphics[scale=0.5,trim={8pt 0 0 0},clip]{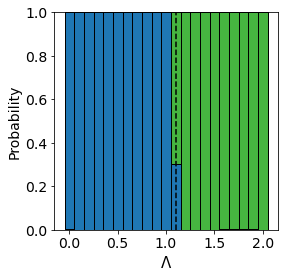} &
            \hspace{-5pt}\includegraphics[scale=0.5,trim={50pt 0 0 0},clip]{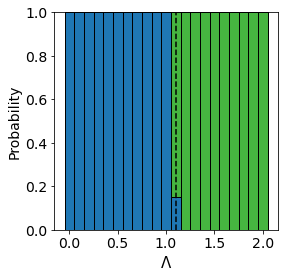}
        \end{tabular} &
        \hspace{-6pt}\begin{tabular}{c c}
            \vspace{-5pt}
            Exact & Trotter \\
            \includegraphics[scale=0.5,trim={8pt 0 0 0},clip]{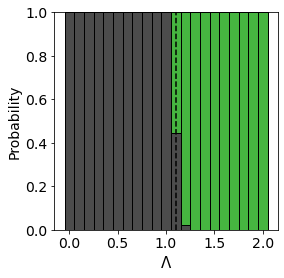} &
            \hspace{-5pt}\includegraphics[scale=0.5,trim={50pt 0 0 0},clip]{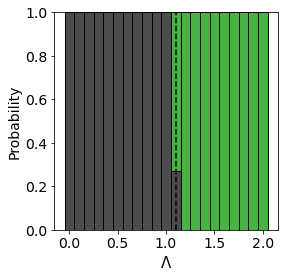}
        \end{tabular} \\
    \end{tabular}
\caption{Quantum phase prediction of the system via the CNN. The graphs show the probability that the system is on a given phase for each point, predicted only from the time evolution of the $C_z(1,2)$ correlation function; both the exact solution and the one obtained from the Trotter expansion with $n_T=6$ are presented for the initial state $|\downarrow_1 \ \downarrow_2 \ \downarrow_3 \ \downarrow_4 \ \uparrow_5 \ \uparrow_6 \ \uparrow_7 \ \uparrow_8 \ \rangle$ with $\varepsilon = 1$. QPTs are shown moving through the following lines of the phase space: (a) $\Sigma = 0.5$ and $\Lambda = 0.5$ moving from $\chi = 0$ (Symmetric) to $\chi = 2$ (HF); (b) $\chi = 0.5$ and $\Lambda = 0.5$ moving from $\Sigma = 0$ (Symmetric) to $\Sigma = 2$ (BCS); (c) $\chi = 0.5$ and $\Sigma = 0.5$ moving from $\Lambda = 0$ (Symmetric) to $\Lambda = 2$ (Combined HF-BCS); (d) $\chi = 1.5$ and $\Lambda = 0.5$ moving from $\Sigma = 0$ (HF) to $\Sigma = 2$ (BCS); (e) $\chi = 1.5$ and $\Sigma = 0.5$ moving from $\Lambda = 0$ (HF) to $\Lambda = 2$ (Combined HF-BCS); (f) $\chi = 0.5$ and $\Sigma = 1.5$ moving from $\Lambda = 0$ (BCS) to $\Lambda = 2$ (Combined HF-BCS). The dashed black line in each graph denotes the theoretical critical point between phases for each case. Reproduced with permission \cite{Saiz2022}. Copyright 2022, American Physical Society.}
\label{fig:Convpred}
\end{figure}

In nuclear models, the shape/phase of the system is determined thanks to  mean-field calculations, although it can also be explored through certain observables that can serve as proxies for QPTs even in finite-size systems \cite{Iach98}. Is it possible to extract information about the phase/shape of the system using a different approach? To answer this question in \cite{perezfernandez2021quantum,Saiz2022} the phase diagram \cite{AgassiPhase1,Garc2018} of the Agassi model \cite{AGASSI196849} has been determined exploring the time evolution of a correlation operator using a quantum computer (quantum simulator) to feed a ML algorithm, i.e., defining a hybrid quantum-classical algorithm. The Agassi model considers a two-level system with $m-$sites in each level. For the fermion operators two indexes are used: $\xi$ labels the level ($+1$ for the upper level and $-1$ for the lower level) and $m$ labels the site within each level. This simple Hamiltonian is interesting because it includes the competition between the monopole-monopole and the pairing interactions and mimics the Kumar-Baranger model for nuclear structure. The Hamiltonian for the extended Agassi model used in \cite{Garc2018} can be written as 
\begin{equation}
H =\varepsilon J^{0}-g\sum_{\xi,\xi^{\prime}=-1,1}^{}A_{\xi}^{\dagger}A_{\xi^{\prime}}  -\frac{V}{2}\left[\left(J^{+}\right)^{2}+\left(J^{-}\right)^{2}\right]-2hA_{0}^{\dagger}A_{0},
\label{H-eAm}
\end{equation}
where the operators in the Hamiltonian are all defined in terms of fermion creation and annihilation operators, $c_{\xi,m}^\dagger$ and $c_{\xi,m}$,
\begin{eqnarray}
{J^{+}\,}&=&{\,\sum_{m=-j}^{j}c_{1,m}^{\dagger}c_{-1,m}\,=\,\left(J^{-}\right)^{\dagger}},\\
{J^{0}\,}&=&{\frac{1}{2}\,\sum_{m=-j}^{j}\left(c_{1,m}^{\dagger}c_{1,m}\,-\,c_{-1,m}^{\dagger}c_{-1,m}\right)},\\
{\,A_{1}^{\dagger}}&=&{\sum_{m=1}^{j}c_{1,m}^{\dagger}c_{1,-m}^{\dagger}} = (A_1)^\dagger,\\
{\,A_{-1}^{\dagger}}&=&{\sum_{m=1}^{j}c_{-1,m}^{\dagger}c_{-1,-m}^{\dagger}} = (A_{-1})^\dagger,\\
\,A_0^\dagger &=& \sum_{m=1}^j \left(c_{-1,m}^{\dagger}c_{1,-m}^{\dagger} - c_{-1,-m}^{\dagger}c_{1,m}^{\dagger} \right) = (A_0)^\dagger.
\end{eqnarray}
%where $c_{\pm1,m}^\dagger$ and $c_{\pm1,m}$ are the fermion creation and annihilation operators, respectively.
\bigskip

The phase diagram for the extended Agassi model is depicted in {\bf Figure \ref{fig-phase-dia-time}a} where the phase transition surfaces are clearly marked together with a pictorial representation of the phases. The phase diagram has been obtained analytically using mean-field techniques \cite{Garc2018}.   
\begin{figure}[htb]
\centering
\includegraphics[width=0.6\linewidth]{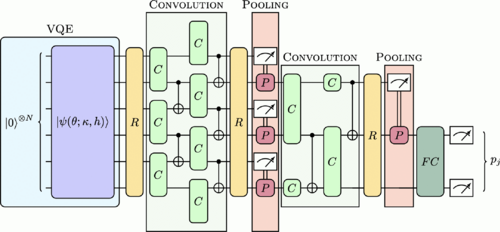}
\caption{Circuit architecture: VQE states (blue) are the input of the quantum CNN composed of free rotations R (yellow), convolutions C (light green), pooling P (red), and a fully connected layer FC (dark green). Reproduced from \cite{Mona2023}. This figure is licensed under CC BY-SA 4.0.}
\label{fig-circuit-QCNN}
\end{figure}

\bigskip

The Hamiltonian (\ref{H-eAm}) can be easily mapped onto a spin Hamiltonian using the Jordan-Wingner (JW) mapping approach \cite{Batista_2001,JordanWigner}, which can later be experimentally implemented into a digital quantum simulator. So far, the simulation has been performed for a system with eight sites ($N=8$, $j=2$). The system is already large from the present quantum computing point of view, but it is really small from the point of view of the QPT analysis, taking into account that the phase of a given system is expected to be properly defined in the large $N$ limit. Let us define the correlation function between two sites, $i, j$ of the system as,
\begin{equation}
    C_{z}(i,j) = 
    \langle \sigma_i^{z} \otimes \, \sigma_j^{z} \rangle -
    \langle \sigma_i^{z} \rangle \, \langle \sigma_j^{z} \rangle,
\end{equation}
where $\sigma_i^{a}$ are the Pauli matrices at site $i$ for $a=x,y,z$. We will consider as an {\it ansatz} that the time evolution of this function with the state $|\downarrow_1 \ \downarrow_2 \ \downarrow_3 \ \downarrow_4 \ \uparrow_5 \ \uparrow_6 \ \uparrow_7 \ \uparrow_8 \ \rangle$ can serve as a proxy to determine the shape of the system and eventually to find the location of the phase transition surfaces. Observing such an evolution with the naked eye one cannot provide any hint about the shape of the system (see panels (b)-(g) of Figure \ref{fig-phase-dia-time} where the time evolution is depicted for selected values of the Hamiltonian parameters). Therefore, the use of a ML technique is in order. In particular, we will focus on the use of a CNN (in \cite{Saiz2022} a Multi-Layer Perceptron is also used). To train the system, a lattice in the control parameter space was created with $9261$ points, reserving the $10\%$ of them for testing purposes. The analysis was performed with the exact evolution operator and also approximating it with a Trotter expansion. The obtained global accuracy was of $98.7\%$ for the exact evolution, while $99.2\%$ for the Trotter one. An appealing fact is that the accuracy of the procedure is even larger when using the Trotter approximation with a small number of steps, which has clear practical advantages. A possible explanation is that the larger oscillations obtained in the approximate evolution, compared with the exact one, gives rise to exaggerated patterns that are easier to recognize. In {\bf Figure \ref{fig:Convpred}}, the results of the CNN analysis are presented for both the exact evolution and the Trotter one for selected values of the Hamiltonian parameters. The reason why the time evolution of a given matrix element is able to describe the phase of the system is that it is connected with the complete spectrum of the Hamiltonian, assuming that the state is not an eigenstate of the Hamiltonian. For instance, a vibrational-like nuclear Hamiltonian will generate a vibrational spectrum while a rotational Hamiltonian will produce a sequence $l(l+1)$. In the first case, the nucleus has a spherical shape, while it is well deformed in the second situation.
\bigskip

Very recently, a similar work, but for the Ising model, has been published \cite{Mona2023}. It shows that the phase diagram in the axial next-nearest-neighbor Ising (ANNNI) model can be obtained using a quantum convolutional neural network (QCNN). The considered Hamiltonian can be written as,
\begin{equation}
    H= \sum^N_{i=1} \Big(\sigma^x_i \sigma^x_{i+1} -\kappa \sigma^x_i  \sigma^x_{i+2}+ h \sigma^z_i \Big ), 
\end{equation}
where $\sigma^a_i$ are the Pauli matrices at position $i$ and the coefficients $\kappa$ and $h$ are taken as positive. The phase diagram of the model is quite rich and three phases are known which are separated by two second-order phase transition lines. The phase diagram of the quantum model at temperature $T = 0$ K  has been studied mainly using the renormalization group or Monte Carlo techniques. To detect the phase in this case, a QCNN was used. The function proposed to train and characterize the phase is the ground state energy of the system. In order to get this energy, a VQE is used and it serves to feed the QCNN. The circuit to perform the process is depicted in {\bf Figure \ref{fig-circuit-QCNN}}.
\begin{figure}[htb]
\centering
\includegraphics[width=0.4\linewidth]{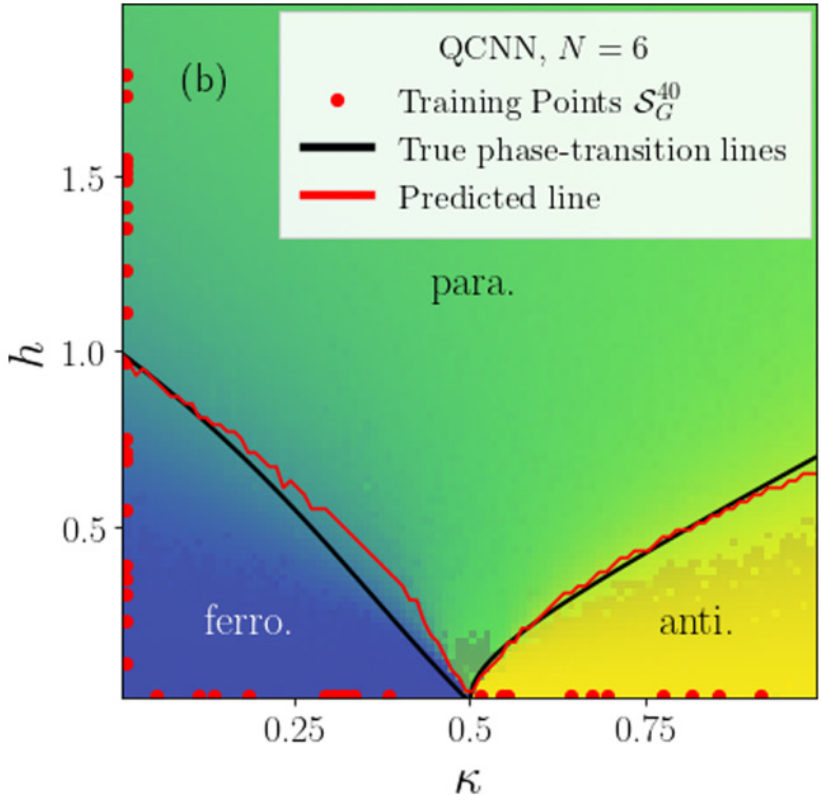}
\caption{Phase diagram of the ANNNI model predicted by the trained QCNN for $N=6$.  Reproduced from \cite{Mona2023}. This figure is licensed under CC BY-SA 4.0.}
\label{fig-ising-phase-dia}
\end{figure}
\bigskip

In order to train the QCNN and fix the variational parameters, %$\theta_i$, 
the cross entropy ${\cal L}$ loss function was used \cite{Mona2023}. The appealing fact of this work is that only very few points were used over the $\kappa=0$ or $h=0$ axes but, nevertheless, the whole phase diagram, including the phase transition lines, was correctly reproduced. In {\bf Figure \ref{fig-ising-phase-dia}} the theoretical phase diagram of the model is depicted superimposing the training points (red dots) and the predicted phase transition lines (red lines). It is really remarkable the ability of the QCCN to disentangle the complete phase diagram including the point with $h=0$ and $\kappa=0.5$ where three phases coexist.  

\subsection{Shell Model Calculations: the Ground State of Nuclear Systems} 
\label{sec-SM}
The determination of the ground state of a nuclear system is one of the central problems of nuclear physics, as it is for quantum chemistry to determine the structure of a given molecule. The exact treatment of this problem is really far from our present knowledge and, consequently, the use of some kind of model is required (see Section \ref{sec-nuclear}). One of these models is the nuclear shell model that provides the ideal starting framework for quantum simulations. Below, different key examples of proposed quantum simulations to calculate ground state energies in nuclear systems are discussed. All of them correspond to the pioneering implementation of the VQE technique to obtain the ground state of an atomic nucleus. The VQE is a variational procedure that strongly depends on the used {\it ansatz}. It is a hard problem to disentangle whether the trial state is the most appropriate or not. To help along this line, it has been recently presented a work in which a reinforcement learning optimization approach is carried out over a variational quantum circuit \cite{Osta2021}. It shows a very remarkable performance in reproducing the ground state energy of the LiH molecule. The implementation of this method in the cases that will be described below within nuclear physics is not simple, but this work clearly proves that the VQE can be optimized using QML.
\begin{figure}[htb]
\centering
\includegraphics[width=0.6\linewidth]{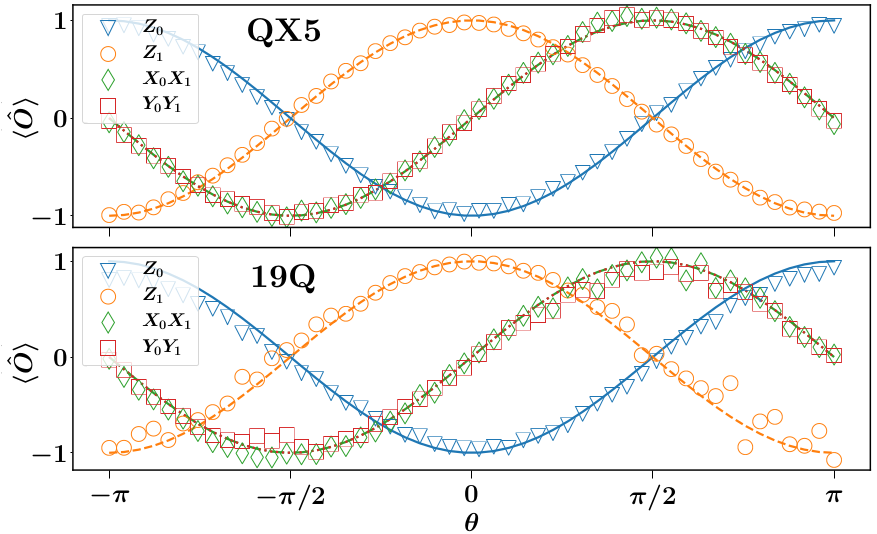}
\caption{Experimentally determined energies for expectation values of the Pauli terms needed to calculate the ground state energy of deuteron as obtained on the QX5 and 19Q chips. Symbols for the experimental results while lines for the theory. Reproduced with permission \cite{Dumi2018}. Copyright 2018, American Physical Society.}
\label{fig-dumi}
\end{figure}

\bigskip

One of the first works along this line was the study of the ground state structure of deuteron in a quantum computer discussed in \cite{Dumi2018}. The deuteron was treated using a Hamiltonian extracted from a pionless effective field theory such that it can be simulated on a quantum chip. The ground state is obtained using a variational wave-function {\it ansatz} based on the unitary coupled-cluster theory (UCC). In the case of the deuteron, the dimension of the Hilbert space is very small and only three single-particle states have been considered. However, a kind of extrapolation can be used and as a result, the energy of the ground state is only $0.5\%$ away from the exact value. The expectation value of the different terms appearing in the Hamiltonian and evaluated in two different quantum computers are presented in {\bf Figure \ref{fig-dumi}}. The $\theta$ variable is the corresponding variational parameter. 
\bigskip

Along the same line, in \cite{Cerv2021} a widely used model that sketches the nuclear interaction, such as the LMG model \cite{LMG}, was implemented in a quantum computer and the ground state was determined using the VQE. In this case, the design of the trial wave function is guided by symmetry considerations of the model and it makes possible to use a single variational parameter, $\theta$, in the wave function for four particles,
\begin{eqnarray}
 |\psi(\theta)\rangle&=&\cos^2 \theta|\downarrow \downarrow \downarrow \downarrow \rangle +
 \sin^2 \theta |\uparrow \uparrow \uparrow \uparrow \rangle \nonumber \\ &+&
 -\frac{1}{\sqrt{12}} \sin 2\theta (|\uparrow \uparrow \downarrow \downarrow\rangle + |\downarrow\downarrow\uparrow\uparrow\rangle + |\downarrow\uparrow\downarrow\uparrow\rangle + |\downarrow\uparrow\uparrow\downarrow\rangle + |\uparrow\downarrow\downarrow\uparrow\rangle + |\uparrow\downarrow\uparrow\downarrow\rangle).
\end{eqnarray}
%In this case, because of the precise algebraic structure of the model, it was possible to tailor a trial wave function that can fully capture the exact ground state, but this is not true in more general systems. 
\bigskip

Also focused in the LMG model, in \cite{Manq2022} the equation of state method, which is an extension of the VQE, is employed to obtain excited states of the system. In this work, two levels of complexity were used, RPA (Random Phase Approximation) or second RPA (SRPA) and they did not find any noticeable difference between both.  
\bigskip

So far, we have seen two types of approaches for defining the variational state, either to use the UCC or to use the symmetry of the Hamiltonian to guide the election of the trial wave function and in both cases were applied to small systems. Next, we will present in detail a set of works where the considered nuclei are heavier and, therefore, their Hilbert spaces are much larger. In these cases, the authors consider as trial wave function the UCC {\it ansatz} with Adaptive derivative-assembled problem-tailored (ADAPT)-VQE \cite{Grims2019} which seems to provide a clear advantage over other VQE approaches. In general, ADAPT-VQE is superior to random or lexical ordering of the excitation operators in terms of convergence and circuit depth. The ADAPT-VQE defines the {\it ansatz} by selecting operators from a pool,
\begin{equation}
    \{\hat \tau_1, \hat\tau_2, \ldots, \hat\tau_N \}
\end{equation}
that present the largest influence on the gradient, i.e., the largest value of
\begin{equation}
    \left|\frac{\partial E}{\partial\theta_i}\right|_{\theta_i=0}=|\langle\psi|[H, \hat\tau_i]|\psi\rangle|.
\end{equation}
Once a new operator from the pool has been selected, $k$ optimization steps are carried out till convergence is reached, before moving into a new term from the pool. It is worth mentioning that with this method the number of trainable parameters does not grow exponentially with the size of the system. The number and type of operators in the pool can be limited thanks to symmetry consideration, which can strongly reduce the complexity of the calculation. This method has been used for $^6$Li \cite{Kiss2022} reaching a precision of a $3.81\%$ for the ground state and $0.12\%$ for the first excited one. The calculation was run in the IBM Quantum 27 qubits architecture \verb+ibmq_mumbai+ using error mitigation. The authors noticed that because the number of nuclear states grows very rapidly with the number of valence nucleons, the scaling of VQE application becomes unfeasible, needing the use of symmetry arguments to reduce the number of operators in the pool. 
\begin{figure}
\centering
\includegraphics[width=0.6\linewidth]{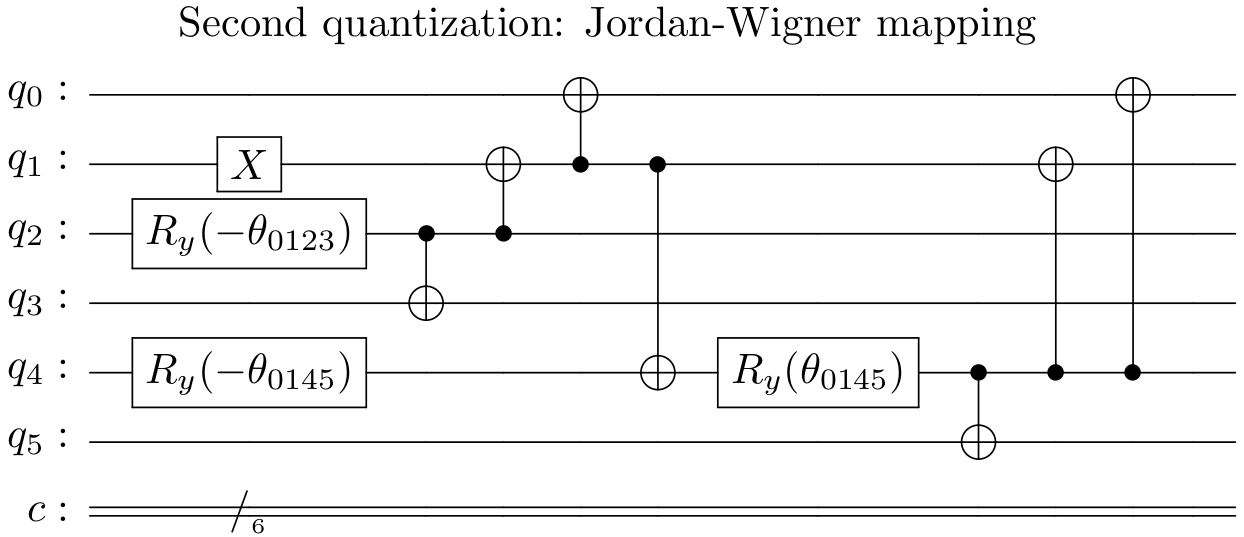}
\caption{Circuit for including a limited number of two-particle, two-hole configurations on top of the Hartree-Fock solution for two particles in 6 states. %, with $R(\theta)=exp(i \theta Y/2)$. For $\theta_{0123}$ and $\theta_{0145}$ equal to zero the Hartree-Fock state is recovered. 
Reproduced with permission \cite{Stet2022}. Copyright 2022, American Physical Society.}
\label{fig-HF-2p-2h}
\end{figure}

\bigskip

A step forward to improve the use of ADAPT-VQE is to start with a more correlated initial state. In general, one can start with a Hartree-Fock state but it is worth exploring the use of other states. In \cite{Stet2022}, the authors use the UCC with ADAPT-VQE, but they include in the initial state two-particle two-hole excitations obtaining a rather good approximation for the ground state energy of $^6$He, $^6$Be, $^{20}$O and $^{22}$O. In {\bf Figure \ref{fig-HF-2p-2h}}, the circuit of a Hartree-Fock state including  two-particle two-hole excitations is shown.
\begin{figure}[hbt]
\centering
\includegraphics[width=0.5\linewidth]{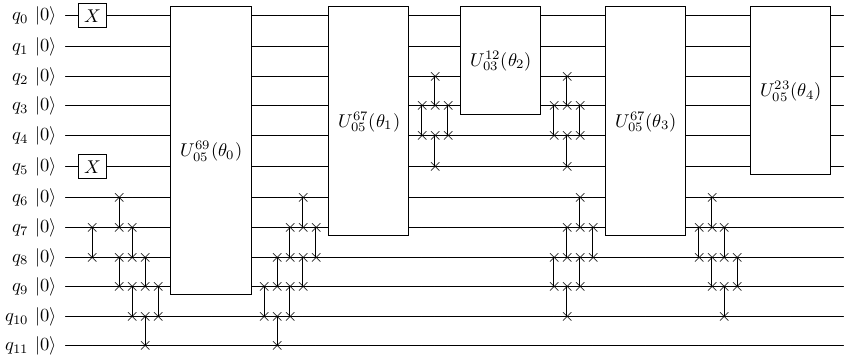}
\caption{Circuit to prepare the $^{18}$O ground state. Reproduced from \cite{Pere2023}. This figure is licensed under CC BY-SA 4.0.}
\label{fig-circuit-O18}
\end{figure}

\bigskip

The latest case to be discussed \cite{Pere2023, Romero2022} also corresponds to the use of ADAPT-VQE to obtain the ground state within the nuclear shell model, but in this case, the authors present a rather general formalism and a large set of potential nuclei can be considered. The formalism is intended for working in the p shell (0p$_{1/2}$, 0p$_{3/2}$ orbitals), the sd shell (1s$_{1/2}$ , 0d$_{3/2}$ and 0d$_{5/2}$ orbitals) or the pf shell (0f$_{7/2}$, 0f$_{5/2}$, 1p$_{3/2}$ and 1p$_{1/2}$ orbitals). The number of single-particle states, either for protons or neutrons, are $6$, $12$, and $20$, respectively. The type of two-body interaction to deal with the above shells is the Cohen-Kurath interaction in the p shell, the USDB in the sd shell and the KB3G interaction in the pf shell. %Explicit three-nucleon interactions can be neglected because their effects can be absorbed into the effective two-body terms \cite{Carb2013}. 
%The Hamiltonian is diagonalized in a suitable many-body basis, such as the M-scheme \cite{Romero2022}. The more popular state-of-the-art nuclear shell-model codes are ANTOINE [57], KSHELL [58], NuShellX [59] and BIG-STICK [60]. All of them use the Lanczos method to obtain iteratively the lowest eigenvalues and eigenvectors. 
The dimension of the Hilbert space will correspond to the product of the dimension of the states for protons and neutrons, which depend in a combinatorial way on the size of the single-particle space and the number of valence particles. This makes unfeasible a direct diagonalization when the dimension is well above of several millions, even using the Lanczos algorithm. Quantum computation has the capability to avoid this problem. The fermion nuclear shell model Hamiltonian is easily mapped into Pauli matrices using a JW transformation. The JW mapping only requires as many qubits as single-particle states, independently of the number of valence particles, which means that the dimension of the problem remains constant for all nuclei belonging to the same major shell. The way to tackle the problem is, once more, the ADAPT-VQE approximation. In this work, the authors explore the complete quantum circuit design to estimate the necessary resources to carry out the nuclear shell-model calculations in regions where the standard approaches cannot be used. In {\bf Figure \ref{fig-circuit-O18}}, it is depicted a quantum circuit with five layers to prepare the ground state of the nucleus $^{18}$O. The multiqubit gates in boxes are defined as $U^{pq}_{rs}(\theta)=e^{i \theta T^{pq}_{rs}}$, where $T^{pq}_{rs}$ is a two-particle promotion operator present in the pool of operators of the ADAPT-VQE method. The state defined in Figure \ref{fig-circuit-O18} has an energy accuracy better than $10^{-6}$.
\bigskip

\begin{figure}
\centering
\includegraphics[width=0.4\linewidth]{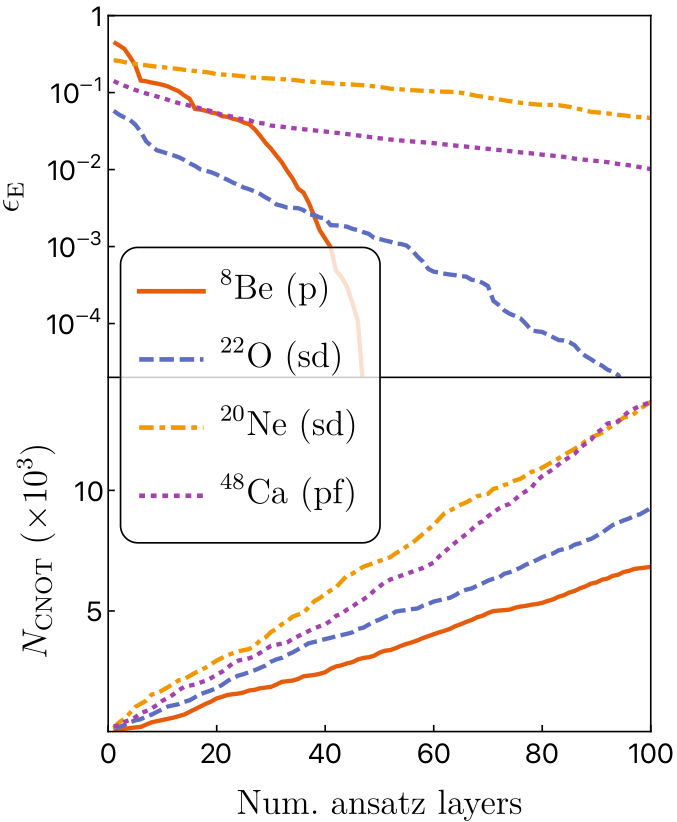}
\caption{Evolution of the relative error for the ground-state energy (top panel) and number of CNOT gates (bottom panel) for selected examples. Reproduced from \cite{Pere2023}. This figure is licensed under CC BY-SA 4.0.}
\label{fig-SM-accuracy}
\end{figure}

An important conclusion of this work is that the accuracy obtained with this approach is increasing faster than the growth of the number of CNOT gates. In {\bf Figure \ref{fig-SM-accuracy}}, the value of the error together with the number of used CNOT gates is depicted and one can easily note that the errors decay exponentially while the number of CNOT gates seems to increase polynomially. The obtained results are very encouraging, having computed the ground state energy for $^6$Be ($10^{-8}$ relative error), $^6$Li, $^{8,10}$Be ($10^{-7}$ relative error), $^{13}$C ($10^{-5}$ relative error), $^{18,19,20, 22}$O ($10^{-6}$ relative error), $^{20,22,24}$Ne ($10^{-2}$ relative error), $^{42}$Ca ($10^{-8}$ relative error) and $^{44,46,48,50}$Ca ($10^{-2}$ relative error). 
\bigskip

Finally, tightly connected with the use of the shell model or other simple models, there are few other publications that deserve to be mentioned. In \cite{Illa2023} the comparison between the use of qubits and qudits is explored in the Agassi model. In  \cite{Robin2023}, the authors studied the effect of using effective model spaces in the quantum computation. The issue of restoring symmetries or preparing states with a given symmetry is analyzed in depth in \cite{Lacr2020,Guzm2022,Guzm2023,Lacr2023}. Finally, in \cite{Rogg2020}, a technique for preparing excited states is presented and in \cite{Rogg2020b},  a neutrino-nucleus scattering calculation has been implemented in a quantum computer. 
%Savage
%\cite{Illa2023}%qudits
%\cite{Robin2023} %Quantum Simulations in Effective Model Spaces (I): Hamiltonian Learning-VQE using Digital Quantum Computers and Application to the Lipkin-Meshkov-Glick Model

%Lacroix
%\cite{Guzm2023}% Restoring broken symmetries using quantum search ``oracles'
%\cite{Lacr2020}% Symmetry assisted preparation of entangled many-body states on a quantum computer
%\cite{Guzm2022} % Accessing ground-state and excited-state energies in a many-body system after symmetry restoration using quantum computers
%\cite{Lacr2023} %Symmetry breaking/symmetry preserving circuits and symmetry restoration on quantum computers

%Stevenson 
%\cite{Hobd2022} %Quantum Computing, Variational Quantum Eigensolver (VQE), Hamiltonian Simulation, Lipkin Model, Nuclear Structure
%\cite{Hobd2023} %Variance minimisation on a quantum computer of the Lipkin-Meshkov-Glick model with three particles

%%Roggero
%\cite{Rogg2020} %Preparation of excited states for nuclear dynamics on a quantum computer
%\cite{Rogg2020b}%Quantum computing for neutrino-nucleus scattering

\subsection{Particle identification and track reconstruction}
\label{sec-track}
The detection and identification of particles, including the measurements of their properties, such as energy, charge, linear or angular momentum is the cornerstone of nuclear and high energy physics experiments since the pioneering works of E.\ Rutherford making collide $\alpha$ particles with a thin foil of gold till nowadays. Nuclear physics and high energy physics experiments present many aspects in common. In particular, the need to know the precise trajectory of the particle inside the detector system or the large number of events to be analyzed (much larger in the case of high energy physics) are common issues of the experiments in both areas of research. 
%Spectroscopy in nuclear physics
In nuclear physics, $\gamma$-ray spectra is a standard technique for isotope identification and fundamental to nuclear structure studies. It is critical to determine the energy spectrum of nuclei and to obtain information about transition probabilities between states, which is essential to disentangle the internal structure of the states. Also, it is of great relevance the spectroscopy of charged particles, $\beta$ or $\alpha$. So far, no QML techniques have been used in nuclear spectroscopy, but classical ML methods have been already used (see Section \ref{sec-nuclear-ML}). %Timing resolution in high-purity germanium detectors has been optimized by fully connected CNN architectures. Another example is the use of CNN for determining the activity in gamma-emitting samples or to identify signal in double-sided silicon strip detectors \cite{Boeh2022}. 
In high energy physics, practical examples of the application QML already exist \cite{Guan2021}, which could serve as inspiration for nuclear physics. Here we describe two of them that could be easily implemented in low-energy nuclear experiments.
\begin{figure}[hbt]
\centering
\includegraphics[width=0.5\linewidth]{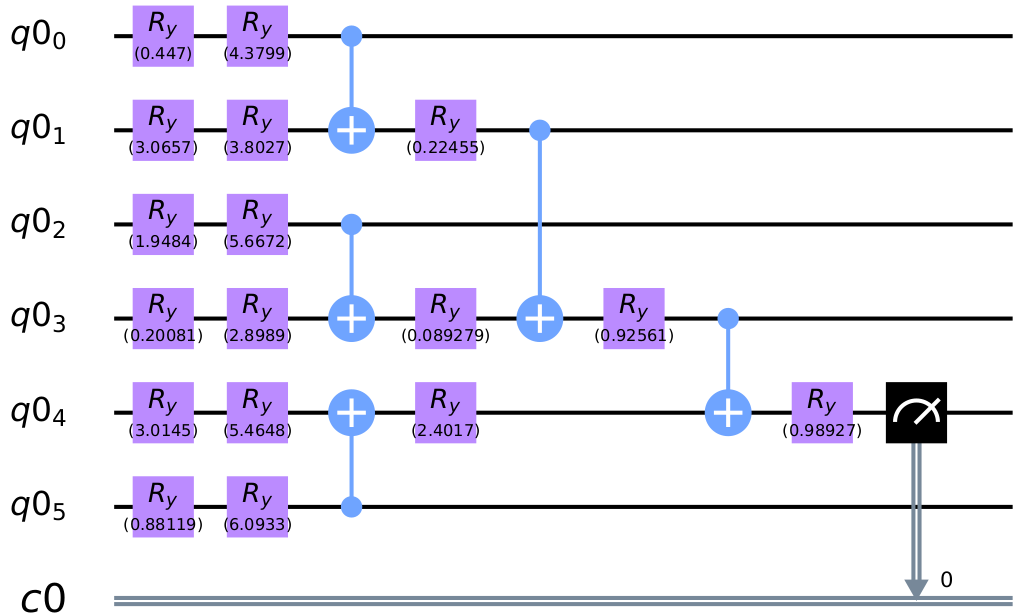}
\caption{Qubit tree tensor network (TTN) representation of the Quantum Edge Network. Reproduced from \cite{Tuys2020}. This figure is licensed under CC BY-SA 4.0.}
\label{fig-track-QCirc}
\end{figure}

\bigskip

%Charge particle tracking \cite{Tuys2020}
The first example corresponds to the determination of the precise trajectory followed by a charged particle or a photon, which is commonly known as {\it tracking}. The {\it tracking} consists in associating to a given particle the hits observed in the system of detectors, and then, reconstructing its trajectory. {\it Tracking} is the cornerstone of particle path reconstruction, which is compulsory to identify the nature of events of interest. In nuclear physics this is a standard technique to increase the efficiency of the detection system, which is especially relevant in experiments with low counting rates. In high luminosity experiments, as happens with LHC experiments, the number of events to be analyzed is really large and, therefore, to carry out the {\it tracking} of particles becomes a challenge. Nowadays, state-of-the-art algorithms are based on the use of Kalman filters. This approach is rather reliable and robust, providing good physics performance. Its main problem is connected with its scalability, which is expected to be worse than quadratically with the increase in the number of simultaneous collisions. Therefore, it is of great interest to explore other approaches to speed up the process including deep machine learning techniques. Such a new avenue is based on the use of image-based interpretation of the detector data where the use of CNN could provide high-accuracy results. The HEPtrkX project \cite{Farr2017} followed this approach and is based on graph neural networks (GNNs) to perform hits and segments classification. The work \cite{Tuys2020} considers a GNN architecture from a quantum computing perspective, implementing  the original networks as quantum circuits.
\bigskip

First, they start with the TrackML dataset, a publicly available dataset that consists of simulated measurements of many detector layers. The detector layers are arranged using a model layout that is common to most LHC experiments. This set of data is first preprocessed prior to the training. The HepTrkX team proposed a GNN  to perform segment classification. The model consists of 3 types of networks. The first one is an input network, the second one is an edge network and the third one is a node network. The model has an overall accuracy of $99.5\%$. 
\bigskip

To transform the GNN into a quantum circuit many modifications are needed. For simplicity, the authors only take into account the edge network and do not use the others networks. Then, it is used the Tree Tensor Network (TTN) among the hierarchical quantum classifiers as the quantum circuit to replace the neural network layer. The input is encoded in qubits and then the TTN circuit is applied. The TTN circuit is made of $R_y$ and CNOT gates. $R_y$ gates start with random parameters that will be tuned later. The CNOT gate is used to introduce correlations between qubits so that their values are not independent. At the end of the circuit, there is a measurement. The structure of the quantum circuit is depicted in {\bf Figure \ref{fig-track-QCirc}}.
\begin{figure}[hbt]
\centering
\includegraphics[width=0.8\linewidth]{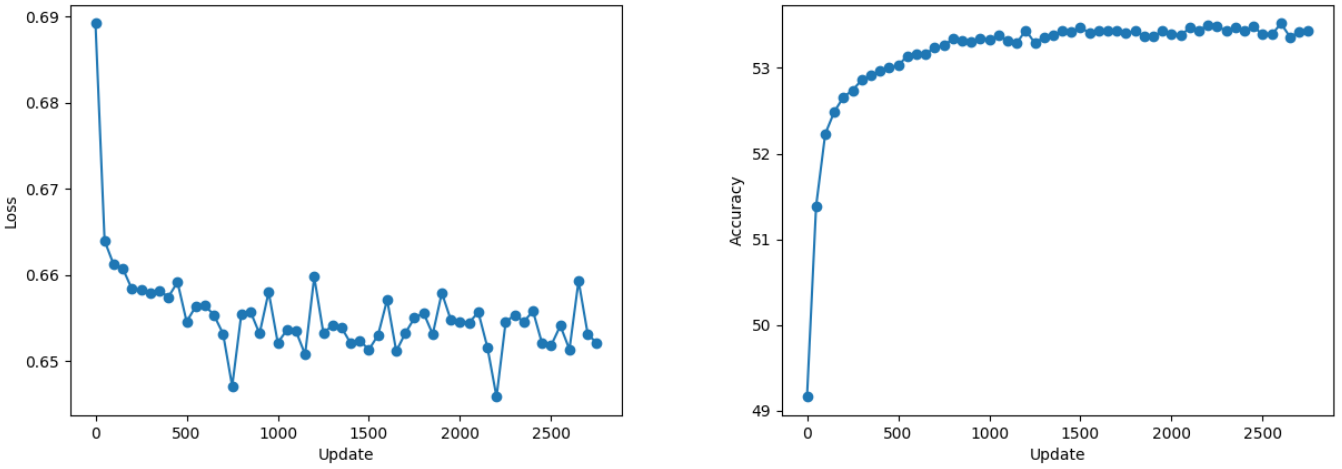}
\caption{Training loss (on the left) and validation accuracy (on the right) of the TTN in 2 full epochs. Reproduced from \cite{Tuys2020}. This figure is licensed under CC BY-SA 4.0.}
\label{fig-track-accuracy}
\end{figure}
The quantum circuit was trained over 2 epochs. The data were divided randomly into training and test sets with a ratio of 9 to 1. The model is trained using stochastic gradient descent and weighted binary cross entropy. The training performance of the model can be seen in {\bf Figure \ref{fig-track-accuracy}}. Note that the obtained accuracy is noticeably small, but this is mainly due to the oversimplification of the model. At the end of the day, this is still a proof of principle prototype of a complete quantum GNN structure.
\bigskip

% b and bar-b identification \cite{Gian2022}
The second example that we will present corresponds to the application of QML to identify if a jet contains a hadron formed by a $b$ (bottom) or $\bar{b}$ quark \cite{Gian2022} at the moment of production. To this end, the Variational Quantum  Classifier is used with simulated data of the LHCb experiment. LHCb is a single-arm spectrometer designed to study $b$ and $c$ (charm) hadrons in the forward region of proton-proton collisions. The goal of this work is to distinguish between jets that contain a $b$ or a $\bar{b}$ hadron just after the hadronization. Therefore the analysis is restricted to a sample of jets that belong to these two categories, either labeled as $b$ jets or as $\bar{b}$ jets. Hence, we have a  binary classification problem. 
\begin{figure}[hbt]
\centering
\includegraphics[width=0.5\linewidth]{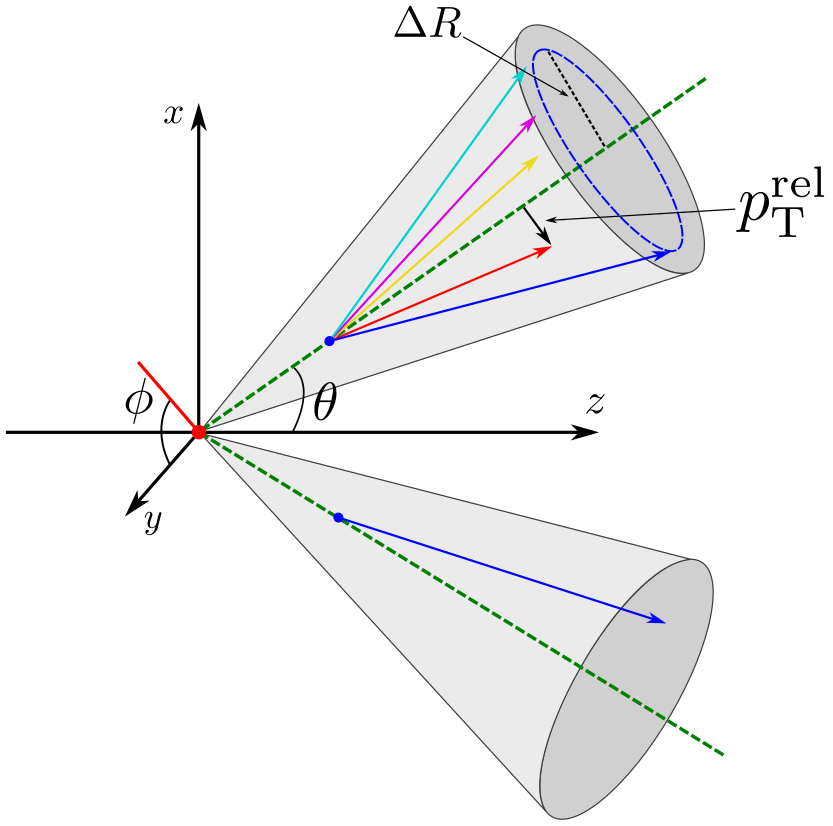}
\caption{Schematic representation of the two jet tagging methods. In the exclusive method the
information comes from a particle, muon (lower jet); in the inclusive method, the information is extracted from the jet constituents (upper jet). Reproduced from \cite{Gian2022}. This figure is licensed under CC BY-SA 4.0.}
\label{fig-bb-cones}
\end{figure}
The QML approach presented in this application belongs to the category of inclusive algorithms (upper jet of {\bf Figure \ref{fig-bb-cones}}). The figure of merit of this work  corresponds to the tagging power, defined as,
\begin{equation}
    \epsilon_{tag}=\epsilon_{eff}(2a-1)^2,
\end{equation}
where $\epsilon_{eff}$ is the tagging efficiency and $a$ the accuracy, i.e. the fraction of correctly tagged jets with respect to the tagged jets.
\bigskip

\begin{figure}[hbt]
\centering
\includegraphics[width=0.8\linewidth]{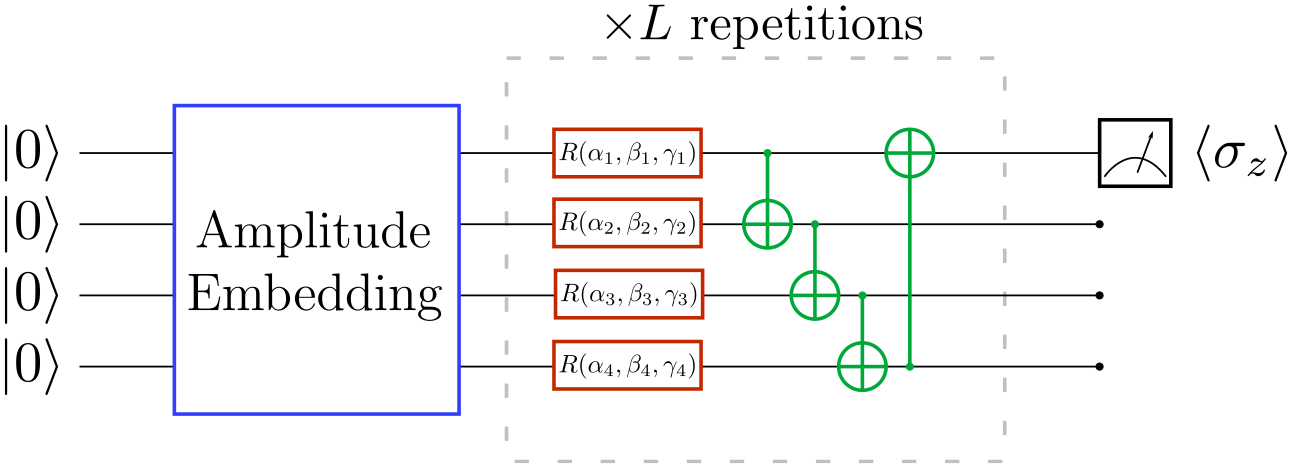}
\caption{Circuit representation of the Amplitude Embedding model. In blue, variables that are
embedded into the amplitudes of a quantum state. In red, trainable gates. In green, CNOT gates entangling qubits with a circular topology. Reproduced from \cite{Gian2022}. This figure is licensed under CC BY-SA 4.0.}
\label{fig-bb-amplitude-emb}
\end{figure}

The QML procedure is implemented with a Variational Quantum Classifier (VQC)  which is a hybrid quantum-classical algorithm to perform classification tasks based on a Parametrized Quantum Circuit (PQC). The structure of the PQC consists in i) the data encoding,  ii) the variational circuit and, iii) the prediction. Two different PQCs are used in this work, the Amplitude Embedding and the Angle Embedding. In {\bf Figure \ref{fig-bb-amplitude-emb}}, we depict the quantum circuit for the first case. The probability of identifying a $b$ or a $\bar{b}$ is connected with the measurement of $\sigma_z$, i.e. $\langle\sigma_z\rangle$. 

\bigskip

In this work, two different datasets are used, namely, the muon and the complete one. As usual, both are split into training and testing sub-datasets: about 60\% of the samples are used in the training and the remaining 40\% are used to test, evaluate and compare the classifiers. 

\bigskip

\begin{figure}[hbt]
\centering
\includegraphics[width=0.8\linewidth]{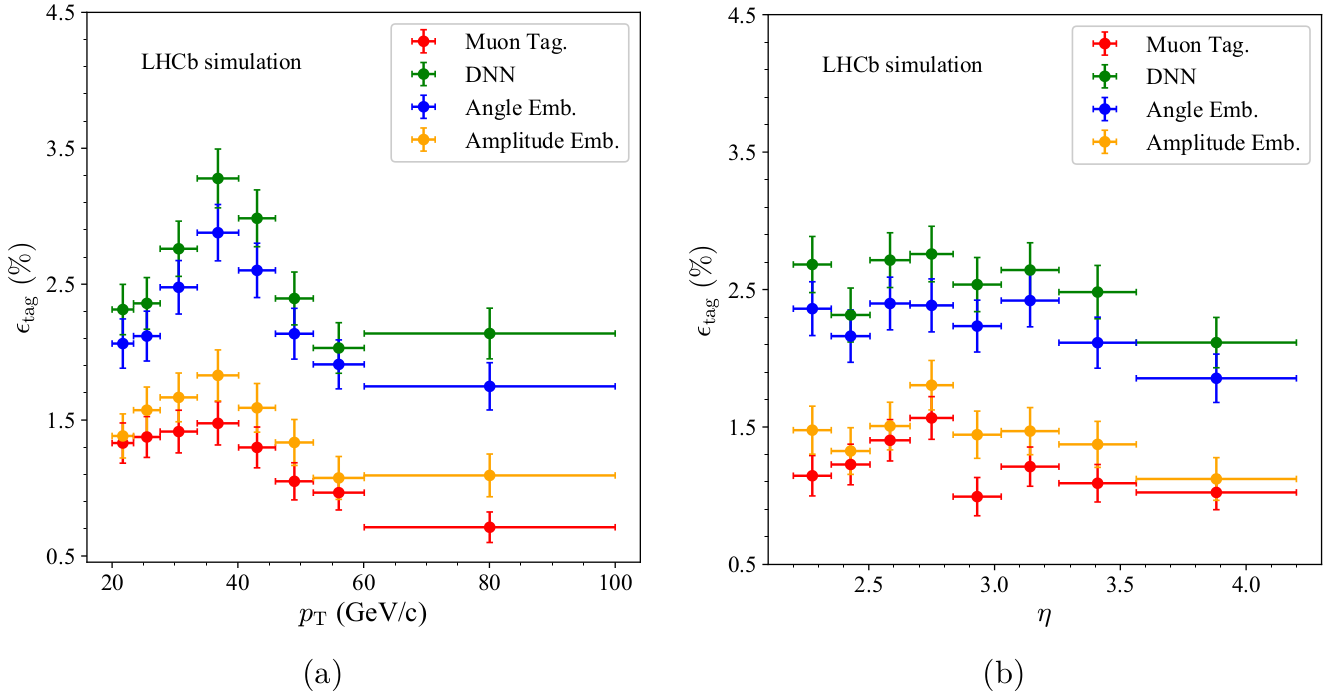}
\caption{Tagging power $\epsilon_{tag}$ with respect to (a) jet $p_T$ and (b) jet $\eta$ for the muon dataset. Reproduced from \cite{Gian2022}. This figure is licensed under CC BY-SA 4.0.}
\label{fig-bb-epsilon-tag}
\end{figure}

In {\bf Figure \ref{fig-bb-epsilon-tag}}, the value of the tagging power as a function of jet p$_T$ and $\eta$ parameters is presented for the classical and quantum classifiers. The results are similar for the different classifiers, without any obvious bias. The tagging power decreases as the p$_T$ value increases because for larger values of $p_T$ the identification is more difficult.  It is worth noting that the deep neural network (DNN) shows slightly better performance compared to the Angle Embedding algorithm, although both values are compatible when considering the error bars. Anyhow, both algorithms reach better results than the Amplitude Embedding model and the muon tagging approach. 

\bigskip

An additional analysis of the value of the tagging power as a function of the number of training events shows that for a large number of events, the performance of the quantum algorithm is similar to the DNN, but when the number of training events decreases, the quantum algorithm keeps very high performance, while the DNN is not able to perform a good classification. Therefore,  with respect to the DNN, the QML method reaches optimal performance with a lower number of events. This special feature of QML algorithms deserves an analysis in depth.

\section{Conclusions and outlook}
\label{sec-conclusions}
Nuclear physics, particularly in the realm of low-energy phenomena, is still in its early stages when it comes to the utilization of quantum computing. While ML techniques have found extensive application in nuclear physics, the application of QML in the context of low-energy nuclear physics is largely unexplored. Consequently, a significant gap exists in the scientific literature regarding this subject.
\bigskip

The objective of this perspective was to provide non-practitioners with essential elements to comprehend ongoing research in nuclear structure studies and to introduce three significant applications of quantum computing and QML in the field of nuclear physics. It is hoped that these examples will inspire future endeavors. The three applications considered were as follows: i) determining the phase and shape of a schematic nuclear system, ii) calculating the ground state of a system described by a shell model Hamiltonian, and iii) determining particle trajectories and identifying particles in nuclear experiments.
\bigskip 

\textbf{Acknowledgements} \par 
This work was partially supported by the Consejer\'{\i}a de Universidad, Investigaci\'on e Innovaci\'on  de la Junta de Andaluc\'{\i}a (Spain) under Groups FQM-160, FQM-177, and FQM-370, and under projects P20-00617, P20-00764,  P20-01247, and US-1380840; by grants PID2019-104002GB-C21, PID2019-104002GB-C22, and PID2020-114687GB-I00 funded by MCIN/AEI/\-10.13039/\-50110001103 and ``ERDF A way of making Europe''.

%\bibliographystyle{MSP}
%\bibliography{references-QC,references-IBM-CM,references-QPT}

\begin{thebibliography}{10}
\providecommand{\url}[1]{\texttt{#1}}
\providecommand{\urlprefix}{URL }

\bibitem{Niel2010}
M.~A. Nielsen, I.~L. Chuang,
\newblock \emph{Quantum Computation and Quantum Information: 10th Anniversary
  Edition},
\newblock Cambridge University Press, \textbf{2010}.

\bibitem{Carl2018}
J.~Carlson, D.~J. Dean, M.~Hjorth-Jensen, D.~Kaplan, J.~Preskill, K.~Roche,
  M.~J. Savage, M.~Troyer,
\newblock {Quantum Computing for Theoretical Nuclear Physics, A White Paper
  prepared for the U.S. Department of Energy, Office of Science, Office of
  Nuclear Physics}, \textbf{2018},
\newblock \urlprefix\url{https://www.osti.gov/biblio/1631143}.

\bibitem{Cloet2019}
I.~C. Cloët, M.~R. Dietrich, J.~Arrington, A.~Bazavov, M.~Bishof, A.~Freese,
  A.~V. Gorshkov, A.~Grassellino, K.~Hafidi, Z.~Jacob, M.~McGuigan, Y.~Meurice,
  Z.-E. Meziani, P.~Mueller, C.~Muschik, J.~Osborn, M.~Otten, P.~Petreczky,
  T.~Polakovic, A.~Poon, R.~Pooser, A.~Roggero, M.~Saffman, B.~VanDevender,
  J.~Zhang, E.~Zohar,
\newblock Opportunities for nuclear physics and quantum information science,
  \textbf{2019}.

\bibitem{Humb2022}
T.~S. Humble, A.~Delgado, R.~Pooser, C.~Seck, R.~Bennink, V.~Leyton-Ortega,
  C.~C.~J. Wang, E.~Dumitrescu, T.~Morris, K.~Hamilton, D.~Lyakh, P.~Date,
  Y.~Wang, N.~A. Peters, K.~J. Evans, M.~Demarteau, A.~McCaskey, T.~Nguyen,
  S.~Clark, M.~Reville, A.~D. Meglio, M.~Grossi, S.~Vallecorsa, K.~Borras,
  K.~Jansen, D.~Krücker,
\newblock Snowmass white paper: Quantum computing systems and software for
  high-energy physics research, \textbf{2022}.

\bibitem{Beck2023}
D.~Beck, J.~Carlson, Z.~Davoudi, J.~Formaggio, S.~Quaglioni, M.~Savage,
  J.~Barata, T.~Bhattacharya, M.~Bishof, I.~Cloet, A.~Delgado, M.~DeMarco,
  C.~Fink, A.~Florio, M.~Francois, D.~Grabowska, S.~Hoogerheide, M.~Huang,
  K.~Ikeda, M.~Illa, K.~Joo, D.~Kharzeev, K.~Kowalski, W.~K. Lai, K.~Leach,
  B.~Loer, I.~Low, J.~Martin, D.~Moore, T.~Mehen, N.~Mueller, J.~Mulligan,
  P.~Mumm, F.~Pederiva, R.~Pisarski, M.~Ploskon, S.~Reddy, G.~Rupak, H.~Singh,
  M.~Singh, I.~Stetcu, J.~Stryker, P.~Szypryt, S.~Valgushev, B.~VanDevender,
  S.~Watkins, C.~Wilson, X.~Yao, A.~Afanasev, A.~B. Balantekin, A.~Baroni,
  R.~Bunker, B.~Chakraborty, I.~Chernyshev, V.~Cirigliano, B.~Clark, S.~K.
  Dhiman, W.~Du, D.~Dutta, R.~Edwards, A.~Flores, A.~Galindo-Uribarri, R.~F.~G.
  Ruiz, V.~Gueorguiev, F.~Guo, E.~Hansen, H.~Hernandez, K.~Hattori, P.~Hauke,
  M.~Hjorth-Jensen, K.~Jankowski, C.~Johnson, D.~Lacroix, D.~Lee, H.-W. Lin,
  X.~Liu, F.~J. Llanes-Estrada, J.~Looney, M.~Lukin, A.~Mercenne, J.~Miller,
  E.~Mottola, B.~Mueller, B.~Nachman, J.~Negele, J.~Orrell, A.~Patwardhan,
  D.~Phillips, S.~Poole, I.~Qualters, M.~Rumore, T.~Schaefer, J.~Scott,
  R.~Singh, J.~Vary, J.-J. Galvez-Viruet, K.~Wendt, H.~Xing, L.~Yang, G.~Young,
  F.~Zhao,
\newblock {Quantum Information Science and Technology for Nuclear Physics.
  Input into U.S. Long-Range Planning, 2023}, \textbf{2023}.

\bibitem{He2020BI}
K.~Heyde,
\newblock \emph{Basic Ideas and Concepts in Nuclear Physics: An Introductory
  Approach},
\newblock Third Edition. Reino Unido, CRC Press, \textbf{2020}.

\bibitem{Ta1993SM}
I.~Talmi,
\newblock \emph{Simple Models of Complex Nuclei},
\newblock New York, Harwood Acad. Publ., \textbf{1993}.

\bibitem{RS2004MB}
P.~Ring, P.~Schuck,
\newblock \emph{The Nuclear Many-Body Problem},
\newblock Berlin, Springer, \textbf{2004}.

\bibitem{Niksic:2011sg}
T.~Niksic, D.~Vretenar, P.~Ring,
\newblock \emph{Prog. Part. Nucl. Phys.} \textbf{2011}, \emph{66} 519.

\bibitem{Grasso:2018pen}
M.~Grasso,
\newblock \emph{Prog. Part. Nucl. Phys.} \textbf{2019}, \emph{106} 256.

\bibitem{RRR-2018-rev}
L.~M. Robledo, T.~R. Rodr\'{\i}guez, R.~R. Rodr\'iguez-Guzm\'an,
\newblock \emph{Journal of Physics G: Nuclear and Particle Physics}
  \textbf{2018}, \emph{46} 013001.

\bibitem{BM1975nuclear}
A.~Bohr, B.~Mottelson,
\newblock \emph{Nuclear Structure, vol 2},
\newblock {W.A. Benjamin, Inc., Reading, Massachusetts, London}, \textbf{1975}.

\bibitem{rowe2010nuclear}
D.~Rowe,
\newblock \emph{Nuclear Collective Motion: Models and Theory},
\newblock World Scientific, \textbf{2010}.

\bibitem{KB1}
K.~Kumar, M.~Baranger,
\newblock \emph{Nuclear Physics A} \textbf{1967}, \emph{92}, 3 608.

\bibitem{KB2}
M.~Baranger, K.~Kumar,
\newblock \emph{Nuclear Physics A} \textbf{1968}, \emph{110}, 3 490.

\bibitem{KB3}
K.~Kumar, M.~Baranger,
\newblock \emph{Nuclear Physics A} \textbf{1968}, \emph{110}, 3 529.

\bibitem{Bohr1998}
A.~Bohr, B.~Mottelson,
\newblock \emph{{Nuclear Structure}},
\newblock World Scientific, \textbf{1998}.

\bibitem{LMG}
H.~Lipkin, N.~Meshkov, A.~Glick,
\newblock \emph{Nuclear Physics} \textbf{1965}, \emph{62}, 2 188.

\bibitem{AGASSI196849}
D.~Agassi,
\newblock \emph{Nuclear Physics A} \textbf{1968}, \emph{116}, 1 49.

\bibitem{Peru014}
A.~Peruzzo, J.~McClean, P.~Shadbolt, M.-H. Yung, X.-Q. Zhou, P.~J. Love,
  A.~Aspuru-Guzik, J.~L. O'Brien,
\newblock \emph{Nature Communications} \textbf{2014}, \emph{5}, 1 4213.

\bibitem{Tilly2022}
J.~Tilly, H.~Chen, S.~Cao, D.~Picozzi, K.~Setia, Y.~Li, E.~Grant, L.~Wossnig,
  I.~Rungger, G.~H. Booth, J.~Tennyson,
\newblock \emph{Physics Reports} \textbf{2022}, \emph{986} 1, the Variational
  Quantum Eigensolver: a review of methods and best practices.

\bibitem{Boeh2022}
A.~Boehnlein, M.~Diefenthaler, N.~Sato, M.~Schram, V.~Ziegler, C.~Fanelli,
  M.~Hjorth-Jensen, T.~Horn, M.~P. Kuchera, D.~Lee, W.~Nazarewicz,
  P.~Ostroumov, K.~Orginos, A.~Poon, X.-N. Wang, A.~Scheinker, M.~S. Smith,
  L.-G. Pang,
\newblock \emph{Rev. Mod. Phys.} \textbf{2022}, \emph{94} 031003.

\bibitem{Geor2014}
I.~M. Georgescu, S.~Ashhab, F.~Nori,
\newblock \emph{Rev. Mod. Phys.} \textbf{2014}, \emph{86} 153.

\bibitem{Bauer2023}
C.~W. Bauer, Z.~Davoudi, N.~Klco, M.~J. Savage,
\newblock \emph{Nature Reviews Physics} \textbf{2023}.

\bibitem{DAQS_Review}
L.~Lamata, A.~Parra-Rodriguez, M.~Sanz, E.~Solano,
\newblock \emph{Advances in Physics: X} \textbf{2018}, \emph{3}, 1 1457981.

\bibitem{QML_Nature_Review}
J.~Biamonte, P.~Wittek, N.~Pancotti, P.~Rebentrost, N.~Wiebe, S.~Lloyd,
\newblock \emph{Nature} \textbf{2017}, \emph{549}, 7671 195.

\bibitem{Lamata_2020}
L.~Lamata,
\newblock \emph{Machine Learning: Science and Technology} \textbf{2020},
  \emph{1}, 3 033002.

\bibitem{Lamata_2023}
L.~Lamata,
\newblock \emph{Advanced Quantum Technologies} \textbf{2023}, 2300059.

\bibitem{Garc2018}
J.~E. Garc\'{\i}a-Ramos, J.~Dukelsky, P.~P\'erez-Fern\'andez, J.~M. Arias,
\newblock \emph{Phys. Rev. C} \textbf{2018}, \emph{97} 054303.

\bibitem{Saiz2022}
A.~S\'aiz, J.~E. Garc\'{\i}a-Ramos, J.~M. Arias, L.~Lamata,
  P.~P\'erez-Fern\'andez,
\newblock \emph{Phys. Rev. C} \textbf{2022}, \emph{106} 064322.

\bibitem{Sach11}
S.~Sachdev,
\newblock \emph{Quantum Phase Transitions},
\newblock Cambridge University Press, Cambridge, UK, \textbf{2011}.

\bibitem{iachello1987interacting}
F.~Iachello, A.~Arima,
\newblock \emph{The Interacting Boson Model},
\newblock Cambridge Monographs on Mathematical Physics. Cambridge University
  Press, \textbf{1987}.

\bibitem{Land69}
L.~Landau, E.~Lifshitz,
\newblock \emph{Statistical Physics},
\newblock Pergamon Press, Oxford, \textbf{1969}.

\bibitem{Iach98}
F.~Iachello, N.~V. Zamfir, R.~F. Casten,
\newblock \emph{Phys. Rev. Lett.} \textbf{1998}, \emph{81} 1191.

\bibitem{perezfernandez2021quantum}
P.~Pérez-Fernández, J.-M. Arias, J.-E. García-Ramos, L.~Lamata,
\newblock \emph{Physics Letters B} \textbf{2022}, \emph{829} 137133.

\bibitem{AgassiPhase1}
E.~D. Davis, W.~D. Heiss,
\newblock \emph{J. Phys. G: Nucl. Phys.} \textbf{1986}, \emph{12}, 9 805.

\bibitem{Mona2023}
S.~Monaco, O.~Kiss, A.~Mandarino, S.~Vallecorsa, M.~Grossi,
\newblock \emph{Phys. Rev. B} \textbf{2023}, \emph{107} L081105.

\bibitem{Batista_2001}
C.~D. Batista, G.~Ortiz,
\newblock \emph{Physical Review Letters} \textbf{2001}, \emph{86}, 6
  1082–1085.

\bibitem{JordanWigner}
P.~Jordan, E.~Wigner,
\newblock \emph{Zeitschrift f\"ur Physik} \textbf{1928}, \emph{47} 631.

\bibitem{Osta2021}
M.~Ostaszewski, L.~M. Trenkwalder, W.~Masarczyk, E.~Scerri, V.~Dunjko,
\newblock In M.~Ranzato, A.~Beygelzimer, Y.~Dauphin, P.~Liang, J.~W. Vaughan,
  editors, \emph{Advances in Neural Information Processing Systems}, volume~34.
  Curran Associates, Inc., \textbf{2021} 18182--18194,
\newblock
  \urlprefix\url{https://proceedings.neurips.cc/paper\_files/paper/2021/\-file/9724412729185d53a2e3e7f889d9f057-Paper.pdf}.

\bibitem{Dumi2018}
E.~F. Dumitrescu, A.~J. McCaskey, G.~Hagen, G.~R. Jansen, T.~D. Morris,
  T.~Papenbrock, R.~C. Pooser, D.~J. Dean, P.~Lougovski,
\newblock \emph{Phys. Rev. Lett.} \textbf{2018}, \emph{120} 210501.

\bibitem{Cerv2021}
M.~J. Cervia, A.~B. Balantekin, S.~N. Coppersmith, C.~W. Johnson, P.~J. Love,
  C.~Poole, K.~Robbins, M.~Saffman,
\newblock \emph{Phys. Rev. C} \textbf{2021}, \emph{104} 024305.

\bibitem{Manq2022}
M.~Q. Hlatshwayo, Y.~Zhang, H.~Wibowo, R.~LaRose, D.~Lacroix, E.~Litvinova,
\newblock \emph{Phys. Rev. C} \textbf{2022}, \emph{106} 024319.

\bibitem{Grims2019}
H.~R. Grimsley, S.~E. Economou, E.~Barnes, N.~J. Mayhall,
\newblock \emph{Nature Communications} \textbf{2019}, \emph{10}, 1 3007.

\bibitem{Kiss2022}
O.~Kiss, M.~Grossi, P.~Lougovski, F.~Sanchez, S.~Vallecorsa, T.~Papenbrock,
\newblock \emph{Phys. Rev. C} \textbf{2022}, \emph{106} 034325.

\bibitem{Stet2022}
I.~Stetcu, A.~Baroni, J.~Carlson,
\newblock \emph{Phys. Rev. C} \textbf{2022}, \emph{105} 064308.

\bibitem{Pere2023}
A.~Pérez-Obiol, A.~M. Romero, J.~Menéndez, A.~Rios, A.~García-Sáez,
  B.~Juliá-Díaz,
\newblock Nuclear shell-model simulation in digital quantum computers,
  \textbf{2023}.

\bibitem{Romero2022}
A.~M. Romero, J.~Engel, H.~L. Tang, S.~E. Economou,
\newblock \emph{Phys. Rev. C} \textbf{2022}, \emph{105} 064317.

\bibitem{Illa2023}
M.~Illa, C.~E.~P. Robin, M.~J. Savage,
\newblock {Quantum Simulations of SO(5) Many-Fermion Systems using Qudits},
  \textbf{2023}.

\bibitem{Robin2023}
C.~E.~P. Robin, M.~J. Savage,
\newblock {Quantum Simulations in Effective Model Spaces (I): Hamiltonian
  Learning-VQE using Digital Quantum Computers and Application to the
  Lipkin-Meshkov-Glick Model}, \textbf{2023}.

\bibitem{Lacr2020}
D.~Lacroix,
\newblock \emph{Phys. Rev. Lett.} \textbf{2020}, \emph{125} 230502.

\bibitem{Guzm2022}
E.~A. Ruiz~Guzman, D.~Lacroix,
\newblock \emph{Phys. Rev. C} \textbf{2022}, \emph{105} 024324.

\bibitem{Guzm2023}
E.~A.~R. Guzman, D.~Lacroix,
\newblock \emph{Phys. Rev. C} \textbf{2023}, \emph{107} 034310.

\bibitem{Lacr2023}
D.~Lacroix, E.~A. Ruiz~Guzman, P.~Siwach,
\newblock \emph{The European Physical Journal A} \textbf{2023}, \emph{59}, 1 3.

\bibitem{Rogg2020}
A.~Roggero, C.~Gu, A.~Baroni, T.~Papenbrock,
\newblock \emph{Phys. Rev. C} \textbf{2020}, \emph{102} 064624.

\bibitem{Rogg2020b}
A.~Roggero, A.~C.~Y. Li, J.~Carlson, R.~Gupta, G.~N. Perdue,
\newblock \emph{Phys. Rev. D} \textbf{2020}, \emph{101} 074038.

\bibitem{Guan2021}
W.~Guan, G.~Perdue, A.~Pesah, M.~Schuld, K.~Terashi, S.~Vallecorsa, J.-R.
  Vlimant,
\newblock \emph{Machine Learning: Science and Technology} \textbf{2021},
  \emph{2}, 1 011003.

\bibitem{Tuys2020}
{T\"uys\"uz, Cenk}, {Carminati, Federico}, {Demirk\"oz, Bilge}, {Dobos,
  Daniel}, {Fracas, Fabio}, {Novotny, Kristiane}, {Potamianos, Karolos},
  {Vallecorsa, Sofia}, {Vlimant, Jean-Roch},
\newblock \emph{EPJ Web Conf.} \textbf{2020}, \emph{245} 09013.

\bibitem{Farr2017}
{Farrell, Steven}, {Anderson, Dustin}, {Calafiura, Paolo}, {Cerati, Giuseppe},
  {Gray, Lindsey}, {Kowalkowski, Jim}, {Mudigonda, Mayur}, {Prabhat},
  {Spentzouris, Panagiotis}, {Spiropoulou, Maria}, {Tsaris, Aristeidis},
  {Vlimant, Jean-Roch}, {Zheng, Stephan},
\newblock \emph{EPJ Web Conf.} \textbf{2017}, \emph{150} 00003.

\bibitem{Gian2022}
A.~Gianelle, P.~Koppenburg, D.~Lucchesi, D.~Nicotra, E.~Rodrigues, L.~Sestini,
  J.~de~Vries, D.~Zuliani,
\newblock \emph{Journal of High Energy Physics} \textbf{2022}, \emph{2022}, 8
  14.

\end{thebibliography}

% \begin{figure}[bht]
%   \includegraphics[width=0.2\linewidth]{figures/photo-je.png}
%   \caption*{Jos\'e-Enrique Garc\'{\i}a-Ramos is full professor of nuclear physics at the Department of Integrated Sciences of the University of Huelva, Spain. Before working in Huelva, he was a postdoctoral fellow at the University of Gent, Belgium. Previously, he carried out his PhD at the University of Sevilla, Spain. Among his research interests, one can highlight the nuclear structure studies using algebraic methods, with emphasis in the nuclear shape-coexistence phenomena, the analysis of quantum phase transitions in nuclear systems and, finally, the use of quantum computing and quantum information concepts to solve nuclear problems.}
% \end{figure}
\end{document}